\documentclass{aa}

\usepackage{natbib}
\usepackage{longtable}
\usepackage{lscape}
\usepackage{amsmath}
\usepackage{txfonts}
\usepackage{color}
\usepackage{url}

\usepackage{graphicx}
\bibpunct{(}{)}{;}{a}{}{,}

\newcommand{\kev}{keV}

\begin{document}
   \title{X-ray properties of NGC~300\thanks{Tables 1 and 2 are also available in electronic form
at the CDS via anonymous ftp to cdsarc.u-strasbg.fr (130.79.128.5)
or via http://cdsweb.u-strasbg.fr/cgi-bin/qcat?J/A+A/}.}
   \subtitle{I. Global properties of X-ray point sources and  their
   optical counterparts}
   \titlerunning{X-ray properties of NGC~300, I.}

   \author{S. Carpano
          \inst{1}
          \and
          J. Wilms
          \inst{2}
          \and
	  M. Schirmer
	  \inst{3,4}
	  \and
          E. Kendziorra
          \inst{1}}

   \offprints{S. Carpano, e-mail: carpano@astro.uni-tuebingen.de}
   \institute{Institut f\"ur Astronomie und Astrophysik, Abteilung
   Astronomie,  Universit\"at T\"ubingen, Sand 1, 72076 T\"ubingen, Germany   
	      \and
   Department of Physics, University of Warwick, Coventry, CV4 7AL,
   United Kingdom
             \and
   Institut f\"ur Astrophysik und Extraterrestrische Forschung,
   Universit\"at Bonn, Auf dem H\"ugel 71, 53121 Bonn, Germany
             \and
   Isaac Newton Group of Telescopes, 38700 Santa Cruz de La Palma, Spain
 }
   \date{Submitted: 13 October 2004; Accepted: 14 July 2005}
   
   \abstract{ We present X-ray properties of NGC 300 point sources,
     extracted from 66\,ksec of \textsl{XMM-Newton} data taken in 2000
     December and 2001 January. A total of 163 sources were
     detected in the energy range of 0.3--6\,\kev. We report on the
     global properties of the sources detected inside the
       $D_{25}$ optical disk, such as the hardness ratio and X-ray
     fluxes, and on the properties of their optical counterparts found
     in B, V, and R images from the 2.2\,m MPG/ESO telescope.
     Furthermore, we cross-correlate the X-ray sources with SIMBAD,
     the USNO-A2.0 catalog, and radio catalogues.  
    }

\maketitle
%

\section{Introduction}
\label{sec:int}

Studies of the X-ray population of spiral galaxies other than our
Galaxy are of importance especially for the understanding of the
formation of X-ray binaries and other X-ray emitting sources.  NGC~300
is a member of the Sculptor galaxy group. Due to its small distance
\citep[$\sim$2.02\,Mpc;][]{freedman}, the SA(s)d dwarf galaxy NGC~300
is an ideal target for the study of the entire X-ray population of a
typical normal quiescent spiral galaxy.  The major axes of the
$D_{25}$ optical disk are 13.3\,kpc and 9.4\,kpc 
\citep[$22'\times 15'$;][]{devau}.  
These studies are even more simplified by the
galaxy's almost face-on orientation and its low Galactic column density 
\citep[$N_\text{H}= 3.6\times 10^{20}\,\text{cm}^{-2}$;][]{dickey:90a}.

The first X-ray population study of \object{NGC~300} was performed
between 1991 and 1997 with a total of five \textsl{ROSAT} pointings
\citep{read2001}.  The total exposure time of these data was 46\,ksec
in the \textsl{ROSAT} Position Sensitive Proportional Counter and
40\,ksec in the \textsl{ROSAT} High Resolution Imager, all with a
nominal pointing position of
$\alpha_\text{J2000.0}=00^\text{h}54^\text{m} 52\fs{}0$ and
$\delta_\text{J2000.0}=-37^\circ 41' 24\farcs 0$.  In these
observations, a total of 29 sources was discovered within the $D_{25}$
disk, the brightest being a black hole candidate with $L_\text{X} =
2.2\times10^{38}\,\text{erg}\,\text{s}^{-1}$ in the 0.1--2.4\,\kev\ 
band.  \citet{read2001} also identified a highly variable
supersoft source and other bright sources coincident with known
supernova remnants (SNRs) and H~\textsc{ii} regions.  The luminosity
of the residual X-ray emission, probably due to unresolved sources and
genuine diffuse gas, has been estimated to  $L_\text{X} =
1.2\times10^{38}\,\text{erg}\,\text{s}^{-1}$ \citep{read2001}.

More recently, NGC~300 was observed with \textsl{XMM-Newton} on 2000
December 26 during \textsl{XMM-Newton}'s revolution 192 and 6\,days
later during revolution 195. Some previous results of these
observations have been presented by \cite{kend} and \cite{carpano}.
Data on the luminous supersoft X-ray source \object{XMMU J005510.7$-$373855} in
the center of NGC~300 were presented by \citet{Kong}.  In addition to
these X-ray data, observations with the 2.2\,m MPG/ESO telescope on La
Silla were performed. Here, we use archival images in the broad band
B, V, and R filters. 

In this paper we report a catalog of the NGC~300 X-ray point sources
obtained with \textsl{XMM-Newton} data, as well as their optical counterparts.
The aim of this work is to present a deeper broad-band catalogue of
X-ray selected sources in NGC~300 to facilitate further population
studies and searches for counterparts in other wavebands.  Detailed
studies of selected X-ray sources will be presented elsewhere (Carpano
et al., in preparation).  The remainder of this work is organized as
follows.  Sect.~\ref{sec:obs} describes the observations and data
reduction of the X-ray and optical data. In Sect.~\ref{sec:prop} we
describe some global properties of the X-ray point sources detected
inside the $D_{25}$ optical disk as well as of NGC 300's central
diffuse region. The analysis of the optical counterparts of the X-ray
sources is presented in Sect.~\ref{sec:optic}. Tables of the X-ray and
optical properties are given in Sect.~\ref{sec:table}. We discuss our
results in Sect.~\ref{sec:conc}.


\section{Observations and data reduction}
\label{sec:obs}
\subsection{X-ray observations and data reduction}
\label{sec:xdata}

\textsl{XMM-Newton} observed NGC~300 during its orbit 192 (2000
December 26; 37\,ksec on source time) and orbit 195 (2001 January 1;
47\,ksec on source time). For both observations, all three EPIC
cameras were operated in their full frame mode with the medium filter.
See \cite{Turner} and \cite{Strueder} for a description of the EPIC
cameras.  The aimpoint of the EPIC-pn camera was centered on NGC~300,
using the same position as that of the earlier \textsl{ROSAT} data.
The good-time-intervals extracted from the MOS light curve for
revolution 192 were also used to filter the events list of the
pn-camera, leaving 30\,ksec of low background data for each of the
three cameras.  The particle background during revolution 195 was low,
resulting in net observing times of 43\,ksec for the two MOS cameras
and 40\,ksec for the pn-camera.

We reduced the data using the standard \textsl{XMM-Newton} Science
Analysis System (SAS), version 6.1.0, using the
\texttt{epchain} task for the EPIC-pn and \texttt{emchain} for the MOS
cameras.  Spectra, images, and lightcurves were extracted using
\texttt{evselect}; we only consider events measured in regions
  away from the CCD borders or bad pixels ($\text{FLAG}=0$), and only
  single and double events for the pn camera ($\text{PATTERN}\le 4$)
  and single to quadruple events for the MOS cameras
  ($\text{PATTERN}\le 12$).  The Response Matrix and Ancilliary
Response files are created with the \texttt{rmfgen} and
\texttt{arfgen} tasks using the newest available calibration files.

\subsection{Optical observations and data reduction}
\label{sec:optdata}

NGC~300 was originally observed between 1999 July and 2000 January
with the 2.2\,m MPG/ESO telescope on La Silla, Chile, for the
ARAUCARIA project \citep{Piet}, an attempt to fine-tune the cosmic
distance ladder by comparing different distance indicators such as
Cepheids, blue supergiants, the tip of the red giant branch, and
planetary nebulae for various nearby galaxies.  The data we used for
this work was retrieved from the ESO archive.  The reduction was
performed in the framework of the Garching-Bonn Deep Survey by
\cite{Schirmer}, who also comment extensively on the data reduction.
NGC~300 was observed throughout 34 nights, which resulted in 11\,hours
(110 images), 10.4\,hours (105 images), and 4.2\,hours (42 images), in
the B, V, and R filters, respectively.  The observations were centered
on $\alpha_\text{J2000.0}=00^\text{h}54^\text{m}50^\text{s}$,
$\delta_\text{J2000.0}=-37^\circ 40' 00''$ with a field of view of
$34'\times 34'$. The average seeing in the B, V, and R data was
$1\farcs 1$, $1\farcs 1$ and $1\farcs 0$, respectively.  The absolute
astrometric accuracy of the optical images is $\sim$0.25\,arcsec. The
relative astrometry accuracy is about ten times better.


\section{Properties of the X-ray detected sources and the diffuse
  emission region}
\label{sec:prop}

\subsection{Source detection}
\label{sec:detec}

Event and attitude file of each instrument were first merged for both
orbits 192 and 195, using the SAS \texttt{merge} task. This
  approach is valid since both observations have the same
  pointing direction and the difference in position angles between the two
  observations was very small and consequently the effect of the
  varying point spread function of \textsl{XMM-Newton} on the
  resulting image is small.  Point source detection was then performed
using a maximum likelihood approach as implemented by the SAS-tool
\texttt{edetect\_chain}. We ran this tool simultaneously on the data
from all three cameras, setting a maximum likelihood threshold of 10
in the 0.3--6.0\,\kev\ band.  After removing sources associated with
the cluster of galaxies CL~0053$-$37, a total of 163 sources were found, of
which 86 sources are within the $D_{25}$ optical disk. As it will be
shown in section~\ref{sec:logn}, our detection limiting flux in the
0.3--6.0\,\kev\ energy band is $F_{0.3-6}\sim 7\times
10^{-16}\,\text{erg}\,\text{cm}^{-2}\,\text{s}^{-1}$ for sources
inside the optical disk.

We adaptively determine source and background regions with the
SAS \texttt{region} task, using an elliptical locus to approximate the
spatially varying point spread function.

\begin{figure*}
 \centering
  \resizebox{\hsize}{!}{\includegraphics[bb=44 173 564 611,clip=true]{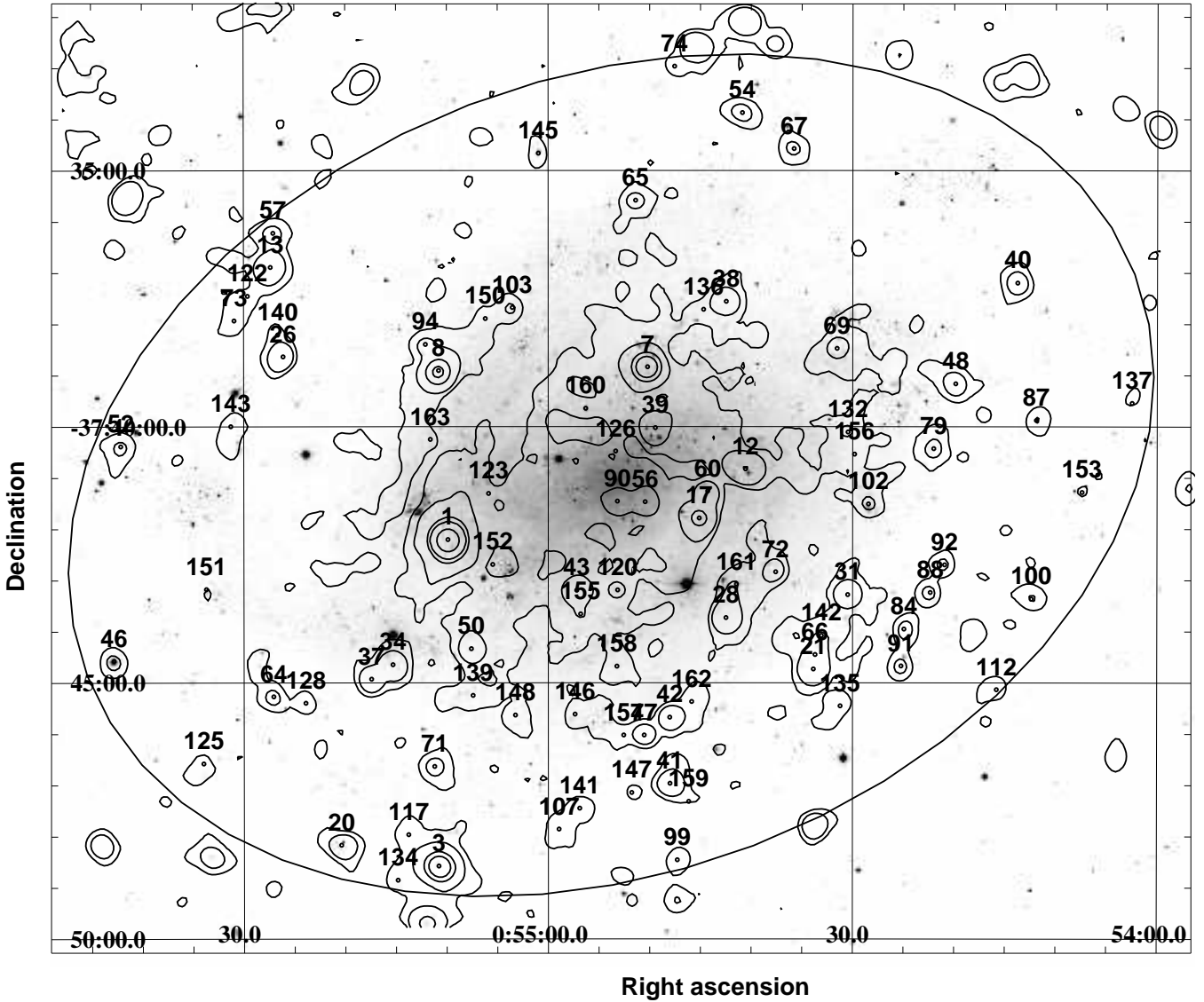}}
 \caption{Optical image of NGC~300 in the visible band overlaid by a
   contour map of the merged 0.3--6.0\,\kev\ raw X-ray image from all
   three EPIC cameras and from both orbits. The $D_{25}$ optical disk
   and the sources detected inside the disk are also shown.}
 \label{fig:sources}
\end{figure*}

Fig.~\ref{fig:sources} shows the V band optical image of NGC~300 and
the contour map of the merged X-ray raw image from both orbits and all
three EPIC cameras in the 0.3--6.0\,\kev\ energy band. The $D_{25}$
optical disk and the sources detected inside the disk, which
are numbered in order of decreasing X-ray count rate as determined by
the \texttt{edetect\_chain}, are also shown.

A summary of the properties of these detected sources as well as their
possible optical counterparts is given in
Table~\ref{tab:table}, described in detail in
Section~\ref{sec:table}.

\subsection{Color-color diagram and X-ray fluxes}
\label{sec:hard}

Any classification of the detected sources as well as the
determination of the source flux require an understanding of the
spectral shape of the sources. Due to the low count rates of most
detected sources, formal spectral modelling is only possible for a few
of the brightest sources (these fits will be shown in a subsequent
paper).  We therefore rely on X-ray color-color and hardness ratio
diagrams in the determination of the flux and the spectral shape.

In order to determine these quantities, we first derive the
background-subtracted count rate from
\begin{equation} 
\mathrm{CR}(I)=\frac{C(I)_\text{src}}{T_\text{src}} - \frac{C(I)_\text{back} B_\text{src}}{T_\text{back}  B_\text{back}},
\end{equation}
where $C(I)$ is the total number of counts in channel $I$, $T$ is the
exposure time and $B$ is the area from which the source and background
data were extracted, as given by the \texttt{BACKSCAL} keyword (see
below). The subscripts `$\text{src}$' and `$\text{back}$' denote the
source and background, respectively.

The \texttt{BACKSCAL} keyword present in the XMM-SAS produced spectra
is defined by the geometric area of the source extraction region minus
the bad pixels or CCD gaps laying within that source region.
Due to software bugs, this keyword is not always correctly
estimated. Source regions intersecting bad CCD columns often have
\texttt{BACKSCAL} overestimated.  For that reason, the total number of
counts in a given energy band (soft, medium, or hard) in
background-subtracted spectra can sometimes be negative.  When this
happens the data coming from that instrument for that revolution are
excluded from the hardness ratio calculation.  To obtain the total
count rate in each band, we add the valid count rates data from all
three EPIC instruments.  The X-ray colors are then defined by:
\begin{equation} 
\text{HR}_\text{hard}= \frac{H-M}{H+M+S},
\quad\text{and}\quad
\text{HR}_\text{soft}= \frac{M-S}{H+M+S}
 \label{equ:hard}
\end{equation}
where $S$, $M$, and $H$ are the total count rates in the soft
(0.3--1.0\,\kev), medium (1.0--2.0\,\kev), and hard (2.0--6.0\,\kev)
energy bands.  The uncertainty of the hardness ratio and the source
countrate is determined by assuming Poisson statistics. Unless
otherwise noted, all uncertainties are at the 68\% level.

\begin{figure}
  \resizebox{\hsize}{!}{\includegraphics[bb=18 11 483 482]{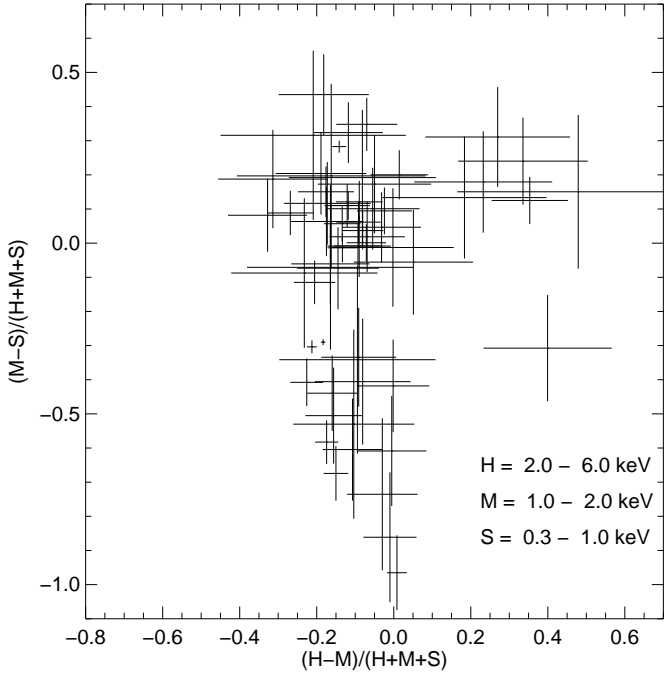}}
 \caption{Color-color diagram of  sources detected inside the
     $D_{25}$ optical disk.} 
 \label{fig:color}
\end{figure}

\begin{figure}
  \resizebox{\hsize}{!}{\includegraphics[bb=18 11 483 482,clip=true]{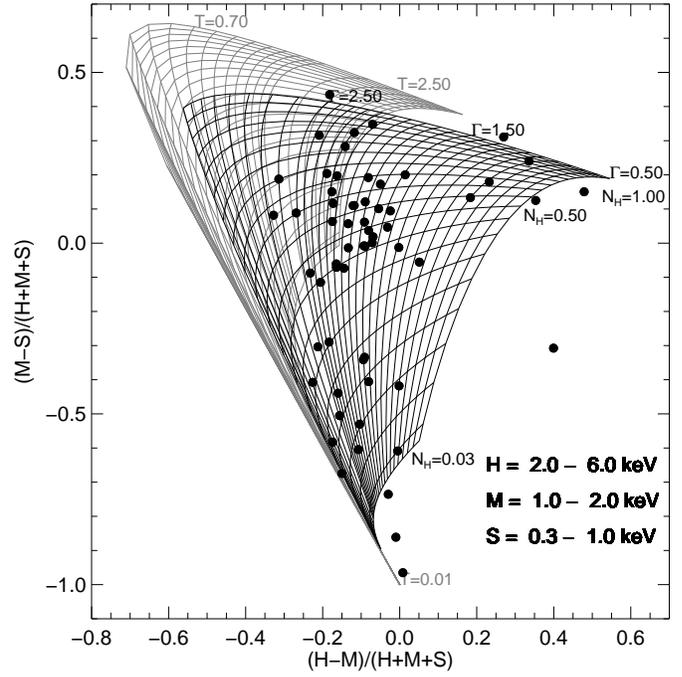}}
 \caption{Color-color diagram of the sources detected inside
     the $D_{25}$ optical disk and color-color contours for
   bremsstrahlung and 0.2\,\kev\ bremsstrahlung plus power law
   component. The equivalent hydrogen column, $N_\mathrm{H}$, is given
   in units of $10^{22}\,\mathrm{cm}^{-2}$, the temperature of the
   bremsstrahlung spectrum, $kT$, is given in \kev.}
 \label{fig:color_fused}
\end{figure}

Fig.~\ref{fig:color} shows the resulting color-color diagram for the
X-ray sources inside the $D_{25}$ optical disk, excluding sources
having less than 20 net counts.  In Fig.~\ref{fig:color_fused} we
compare these data with empirical color-color diagrams assuming a
simple bremsstrahlung model and a two component source spectrum
consisting of a soft bremsstrahlung and a hard power law component
(colors derived from simple power law models were not sufficient to
describe the data).  In these models the equivalent hydrogen column
$N_\mathrm{H}$, expressed in units of $10^{22}\,\mathrm{cm}^{-2}$, is
running from 0.03 to 1.0. In the simple bremsstrahlung model the
temperature $kT$ varies from 0.01 to 5.0\,keV. In the bremsstrahlung
plus power-law model the photon index $\Gamma$ varies from 0.5 to 4.5
and the bremsstrahlung temperature is fixed at 0.2\,\kev.  Both models
are sufficient to describe the data, however, the $N_\text{H}$ values
inferred are generally larger than the pure Galactic $N_\text{H}$ in
the direction to NGC~300 \citep[which is $3.6\times 10^{20}\,\text{cm}^{-2}$;][]{dickey:90a}, 
indicating intrinsic
absorption within NGC~300 and also pointing towards a possible
contamination of our source sample by background AGN. From the
2--10\,keV AGN $\log N$--$\log S$-diagram of \citet{ueda:03a},
$\sim$30 AGN with
$F_{2-10}\ge10^{-14}\,\text{erg}\,\text{cm}^{-2}\,\text{s}^{-1}$ are
expected within the $D_{25}$-disk, however, the identification of AGN
in our sample requires X-ray spectral analysis which is only possible
for the brightest sources and dangerous in itself due to the
similarity of AGN and XRB spectra.

Our color-color diagram analysis shows that for all sources except one
it is possible to find a best matching bremsstrahlung or
bremsstrahlung plus power law model. The spectrum of the one
non-matching source (\#120), which is in a very complex region, 
is peculiar and has been excluded from
the subsequent analysis.  The low number of counts precludes any
further statement about the nature of this source.

From this best matching spectral model it is then possible to
determine the flux of a source by appropriately scaling the flux
determined from the spectral model to the source count rate.  The
uncertainty of the flux is derived from the minimum and maximum value
of fluxes as determined from the error box of the color-color space
defined by the source colors.  Note that such an approach gives
generally more believable flux estimates than the more commonly used
approach of assuming one fixed spectral shape for all detected
sources, while not limiting one to determining spectral fluxes only
for sources with sufficient counts to enable formal spectral model
fitting \citep[see also][]{humphrey:04a}.

\begin{figure}
 \resizebox{\hsize}{!}{\includegraphics[bb=0 0 500 490,clip=true]{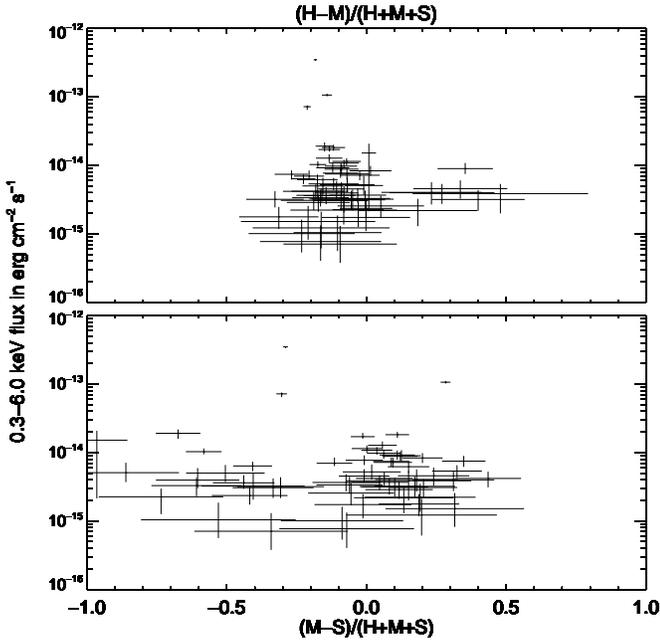}}
 \caption{Fluxes of the sources detected inside the $D_{25}$
     optical disk as a function of the harder (top) and softer
   (bottom) hardness ratio defined by equation (\ref{equ:hard}).}
 \label{fig:flux}
\end{figure}

Fig.~\ref{fig:flux} shows the source fluxes versus both hardness ratios
defined by Eq.~\eqref{equ:hard}.  For comparison, a source with a
luminosity of $1.82\times10^{38}\,\text{erg}\,\text{s}^{-1}$, close to
the Eddington limit for a $1.4\,M_\odot$ object, has an integral flux
of $3.73\times10^{-13}\,\text{erg}\,\text{cm}^{-2}\,\text{s}^{-1}$ at
the distance of NGC~300.

Our brightest source, source 1, is found with a luminosity of
  $1.70\times10^{38}\,\text{erg}\,\text{s}^{-1}$ which is very close to
the Eddington limit for a $1.4\,M_\odot$ object.  The source, coincident
  with the previously known \textsl{ROSAT} source P42
\citep{read2001}, has been found to have a slightly lower
luminosity than in the \textsl{ROSAT} observations, but spectral
fitting of the brightest X-ray sources, which will be given in a
forthcoming paper, is needed to certify if these sources have an
intrinsic variability.  From their high intrinsic luminosities, these
sources are akin to $\sim$$10\,M_\odot$ black holes in their soft
state such as LMC~X-1 or LMC~X-3 \citep{nowak:01a,wilms:99d}. There
are no clear super-Eddington X-ray sources detected in NGC~300.
Finally, we also note that both hardness ratios do not depend
significantly on flux.

\begin{figure}
 \resizebox{\hsize}{!}{\includegraphics[bb=5 4 500 500,clip=true]{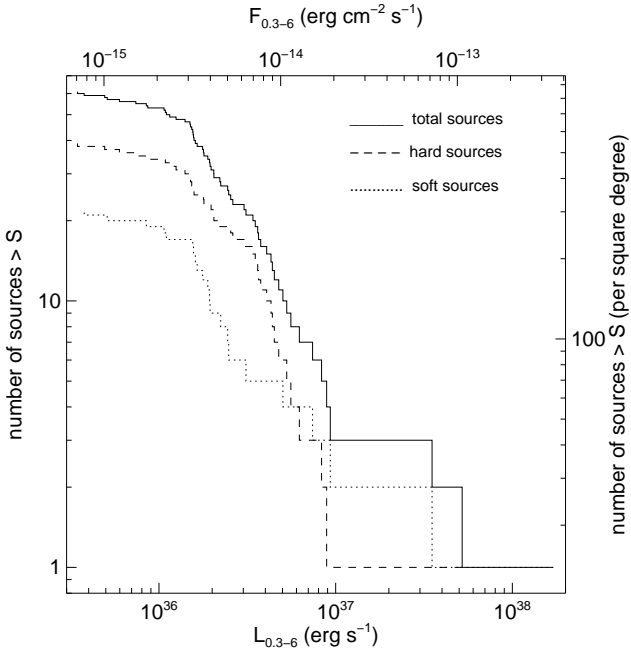}}
  \caption{$\log N$--$\log S$ diagram of all sources 
    with more than 20 net counts inside the optical
    $D_{25}$ disk (solid line), of soft sources with
    $\text{HR}_\text{soft} \leq -0.2$ (dotted line), and for sources
    harder than that (dashed line). }
 \label{fig:logNlogS}
\end{figure}

\subsection{The luminosity function of NGC~300}
\label{sec:logn}
Fig.~\ref{fig:logNlogS} shows the $\log N$--$\log S$ diagram for all
detected sources that are inside the optical disk of NGC~300
and having more than 20 net counts, expressed as a function
of their X-ray luminosity and flux.  Note that we do not make an
attempt to correct for possible background AGN, which could appear as
sources which are strongly absorbed by the gas within NGC~300.

The break of the power law at a luminosity of $1.5\times
10^{36}\,\text{erg}\,\text{s}^{-1}$ (corresponding to $F_{0.3-6}\sim
3\times 10^{-15}\,\text{erg}\,\text{cm}^{-2}\,\text{s}^{-1}$) defines
our completeness limit. Describing the luminosity function above this
limit by a pure powerlaw, $N\propto L^{-\alpha}$, we use 
a Maximum-Likelihood method in the form suggested by \cite{Crawford1970}.
We find a slope of $\alpha=1.17\pm 0.17$ (since our source sample is lacking
objects with $L_\text{X}\gg 10^{38}\,\text{erg}\,\text{s}^{-1}$, such
a simple power law \textsl{Ansatz} for the luminosity function is
sufficient; see, e.g., \citealt{humphrey:04a}).  This slope of the
NGC~300 luminosity function is similar to the slope of the disk
population in several other nearby spirals such as \object{M~31}
\citep[$\alpha=0.9\pm 0.1$;][]{williams:04a} or \object{NGC~1332}
\citep[][]{humphrey:04a}, and also in agreement with the mean slope
for nearby spiral galaxies 
\citep[$\alpha=0.79\pm 0.24$;][]{colbert:04a}.

Due to the apparent bimodality of the sources in Fig.~\ref{fig:color},
we define two subclasses of sources: \emph{hard sources}, defined by
$\text{HR}_\text{soft} > -0.2$, and \emph{soft sources}, with
$\text{HR}_\text{soft}\le -0.2$.  Fitting a pure power law to
  both curves, we find a slope of $1.12\pm 0.27$ and $1.23\pm 0.22$
  for the soft and the hard sources respectively.  Excluding sources
  above a limiting luminosity of $1\times
  10^{37}\,\text{erg}\,\text{s}^{-1}$ (excluding two sources in the
  soft and one in the hard sample), the Maximum-Likelihood method gives a
  slope of $\alpha_\text{soft}=2.03\pm 0.52$ and
  $\alpha_\text{hard}=1.39\pm 0.26$ for the soft and the hard sources
  respectively.  The soft power law slope found here is a bit higher than
  that of the Milky Way HMXB \citep[][finding a slope of
  $\alpha=0.6^{+0.14}_{-0.12}$]{Grimm}. The shape of the hard sources
  is instead more complex.  \citet{Grimm} described the Milky Way LMXB
  luminosity function by a modified power law which takes into account
  the gradual steepening of the $\log N$-$\log S$ relation towards
  higher fluxes. There are indications that the hard sources in NGC~300
  follow a similar luminosity function, as is indicated by the
  different slope for sources with luminosities between $1.5\times
  10^{36}\,\text{erg}\,\text{s}^{-1}$ and $\sim3.5\times
  10^{36}\,\text{erg}\,\text{s}^{-1}$, and for sources between
  $\sim3.5\times 10^{36}\,\text{erg}\,\text{s}^{-1}$ and $1\times
  10^{37}\,\text{erg}\,\text{s}^{-1}$. Due to the low number of
  sources in our sample, however, constraining the luminosity function
  in this range is not possible.  We also note that the two
luminosity functions cross at $\sim$$6\times
10^{36}\,\text{erg}\,\text{s}^{-1}$ and that it is the soft sources
which are dominating at the highest luminosity levels, as seems to be
typical for spiral galaxies \citep{colbert:04a}.

\subsection{The central diffuse emission region}
\label{sec:diff}
\begin{figure}
  \centering
   \resizebox{\hsize}{!}{\includegraphics{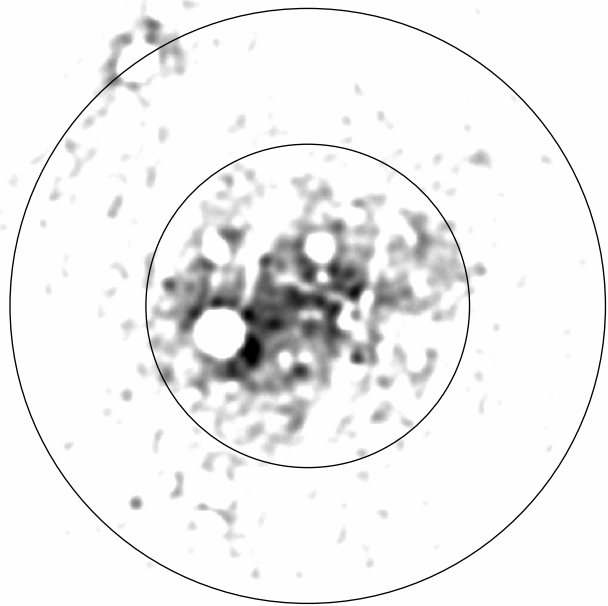}}
  \caption{Smoothed 0.3--1.3\,\kev\ X-ray image of
    the central region of NGC~300 after removal of detected sources.
    The circle and annulus show the region for the central diffuse
    emission area and the associated background, respectively.}
  \label{fig:diff}
 \end{figure}

We extracted the spectrum of the central diffuse emission region after
removing all point sources located in that region. The
  extracted regions are defined by the \texttt{region} task, such that
  the brightness contour level of the source PSF functions are equal
  to half of their background flux.  We defined the diffuse emission
region with a circle of radius $386\farcs 5$, centered on
$\alpha_\text{J2000.0}=00^\text{h}54^\text{m}52\fs{}4$,
$\delta_\text{J2000.0}=-37^\circ41'07\farcs 3$. The background was
taken from an annulus with the same center, an inner radius of
$386\farcs 5$ and outer radius of $711\farcs 1$ (see
Fig.~\ref{fig:diff}).

\begin{figure}
  \centering \resizebox{\hsize}{!}{\includegraphics[bb=115 44 563
    708,clip=true,height=6cm,width=6cm,angle=-90]{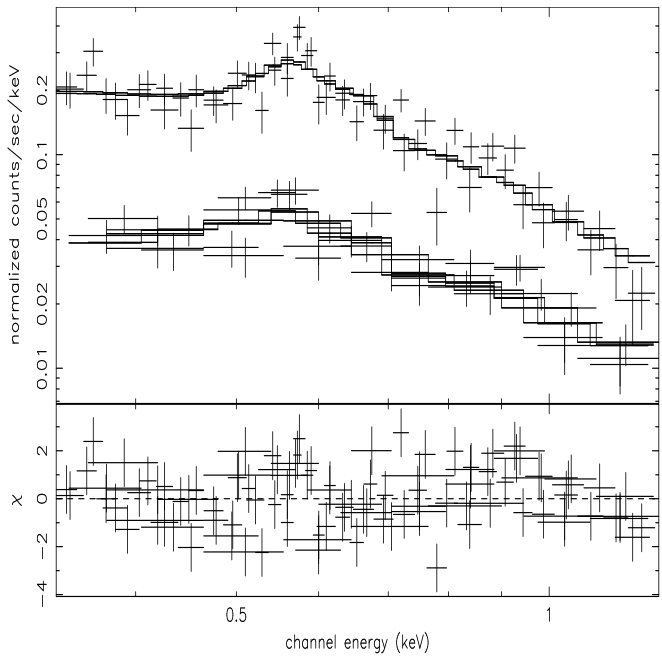}}
  \caption{top: EPIC pn and MOS spectra of the
    central diffuse emission region and the best fit spectral model,
    consisting of the sum of an APEC model,  and a
      bremsstrahlung component, bottom: residuals expressed in
    $\sigma$.}
  \label{fig:apec}
 \end{figure}

Data from all instruments and both revolutions were used to extract
the spectrum.  Because the spectrum of the diffuse emission
  region is very soft, we consider only the 0.3--1.3\,\kev\ energy
  band. The Al and Si fluorescence lines present in the MOS
  background (in the 1.3--1.9\,\kev\, band), which cannot be removed properly, are beyond the region
  of interest.  The spectrum can be described
  ($\chi^2/\text{dof}=142.5/94$) by thermal emission from a
  collisionally ionized plasma, as described by XSPEC's APEC model
  (see \texttt{http://hea-www.harvard.edu/APEC/} for a description of
  this model) with a temperature of $kT=0.2\pm 0.01$\,\kev\ 
  plus a thermal component with a temperature of
  $kT=0.8\pm 0.1$\,\kev.  The 0.3--1.3\,\kev\, flux is
  $F_\text{0.3--1.3}=1.8\times10^{-13}\,\text{erg}\,\text{cm}^{-2}\,\text{s}^{-1}$
  (Fig.~\ref{fig:apec}; error bars are at the 90\% level).  Similar
results are found for the diffuse region in nearly face-on spiral
galaxy M101 \citep{Kuntz}, where the spectrum in the 0.5--2\,\kev\ 
band, is characterized by the sum of two thermal spectra with
$kT=0.20$\,keV\ and $kT=0.75$\,keV.

\section{Properties of the optical counterparts}
\label{sec:optic}

To determine the optical counterparts of the X-ray sources we first
improved the X-ray aspect solution by comparing the optical and X-ray
coordinates of 21 sources inside of the $D_{25}$ disk which have clear
optical counterparts. This is done via the \texttt{eposcorr} task,
which uses a correlation algorithm to find offsets in RA, DEC, and
roll angle which improve the positional accuracy of the X-ray
positions with respect to the optical data.  These optimum offsets are
then used to correct the input source positions.  This algorithm
reveals a systematic shift (X-ray $-$ optical) of $-1.25''\pm 0.31$ in
right ascension, of $-0.17''\pm 0.31$ in declination, and of
$-0\fdg{}077\pm 0\fdg{}27$ for the roll angle.

These offset values are close to values found in the astrometric
calibration of \textsl{XMM-Newton} by
\citet{guainazzi}\footnote{\url{http://xmm.vilspa.esa.es/external/xmm_sw_cal/sas_sci_val/index.shtml}},
who find $-2.52''$ ($1\sigma$) and $-3.09''$ ($1\sigma$) in right
ascension for MOS1 and MOS2, and $1\farcs{}19$ and $0\farcs{}41$ in
declination, respectively.  The final uncertainty in X-ray position
results from a combination of the \texttt{edetect\_chain} output and
the error associated with this position offset.

After correcting the X-ray positions, we searched for all
possible optical counterparts in the merged BVR optical image and then
calculate their fluxes in each of these three optical bands.
Photometry was performed with the IDL \texttt{idlphot} photometry
library available at
\url{http://idlastro.gsfc.nasa.gov/contents.html}, which is a set of
IDL procedures adapted from an early Fortran version of the DAOPHOT
aperture photometry package \citep{stetson}.

We generate an initial optical catalogue by searching for sources
within the area surrounding the corrected X-ray positions (for which
the radius is given by the uncertainty of the position) in the merged
optical image using \texttt{idlphot}'s \texttt{find} procedure and
assuming a Gaussian point spread function (PSF).  This search results
in a list of several possible optical counterparts.  These source
positions are then improved by fitting a measured PSF (as determined
from bright optical sources in the image) and the source flux in the
B, V, and R bands is determined from the PSF fit after subtracting the
local background level. Comparing the B and V magnitudes with
reference stars given by \cite{piet2002} shows differences of less
than 0.15\,mag, in agreement with our typical flux uncertainty.

\begin{figure*}
\centering
  \resizebox{12cm}{!}{\includegraphics{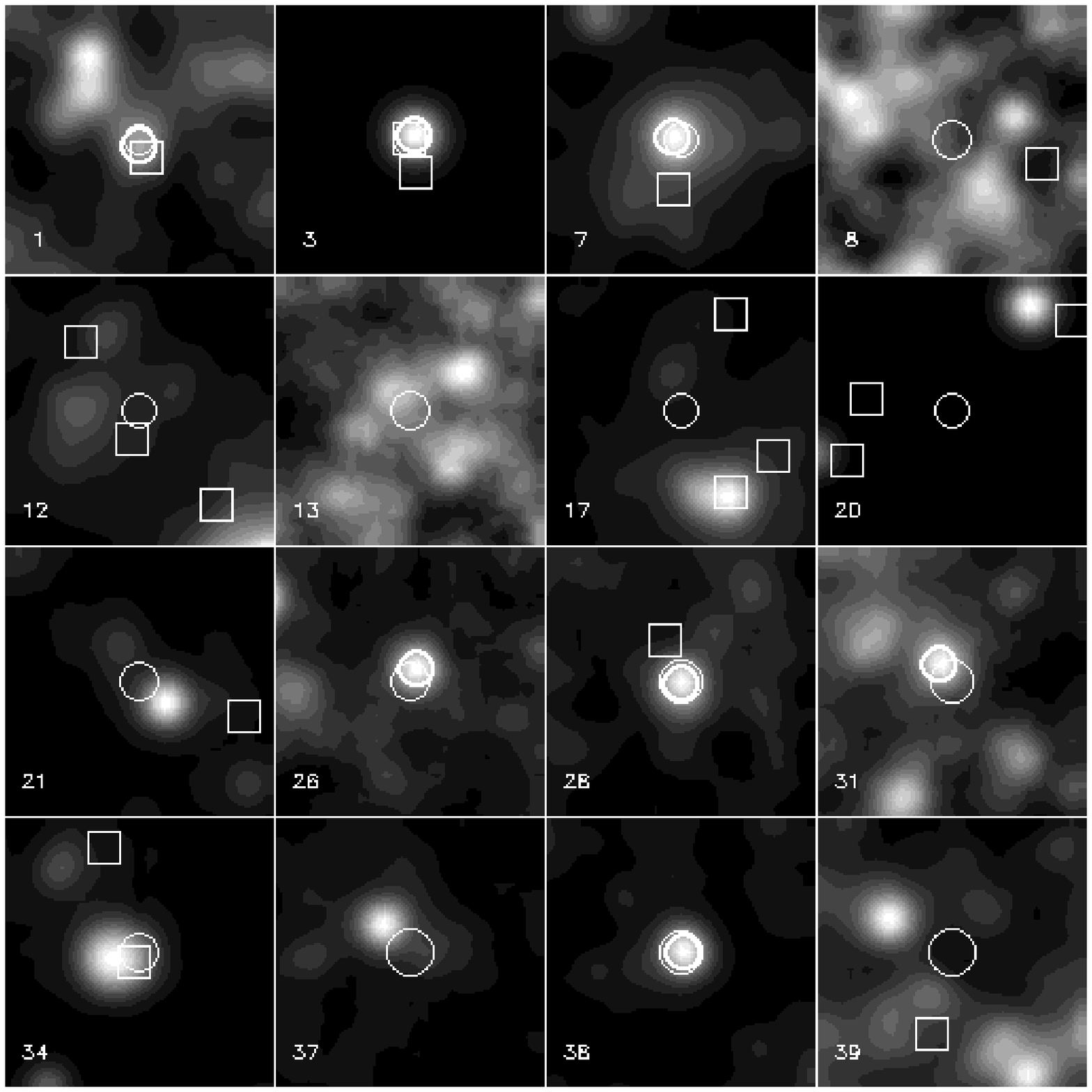}} 
  \resizebox{12cm}{!}{\includegraphics{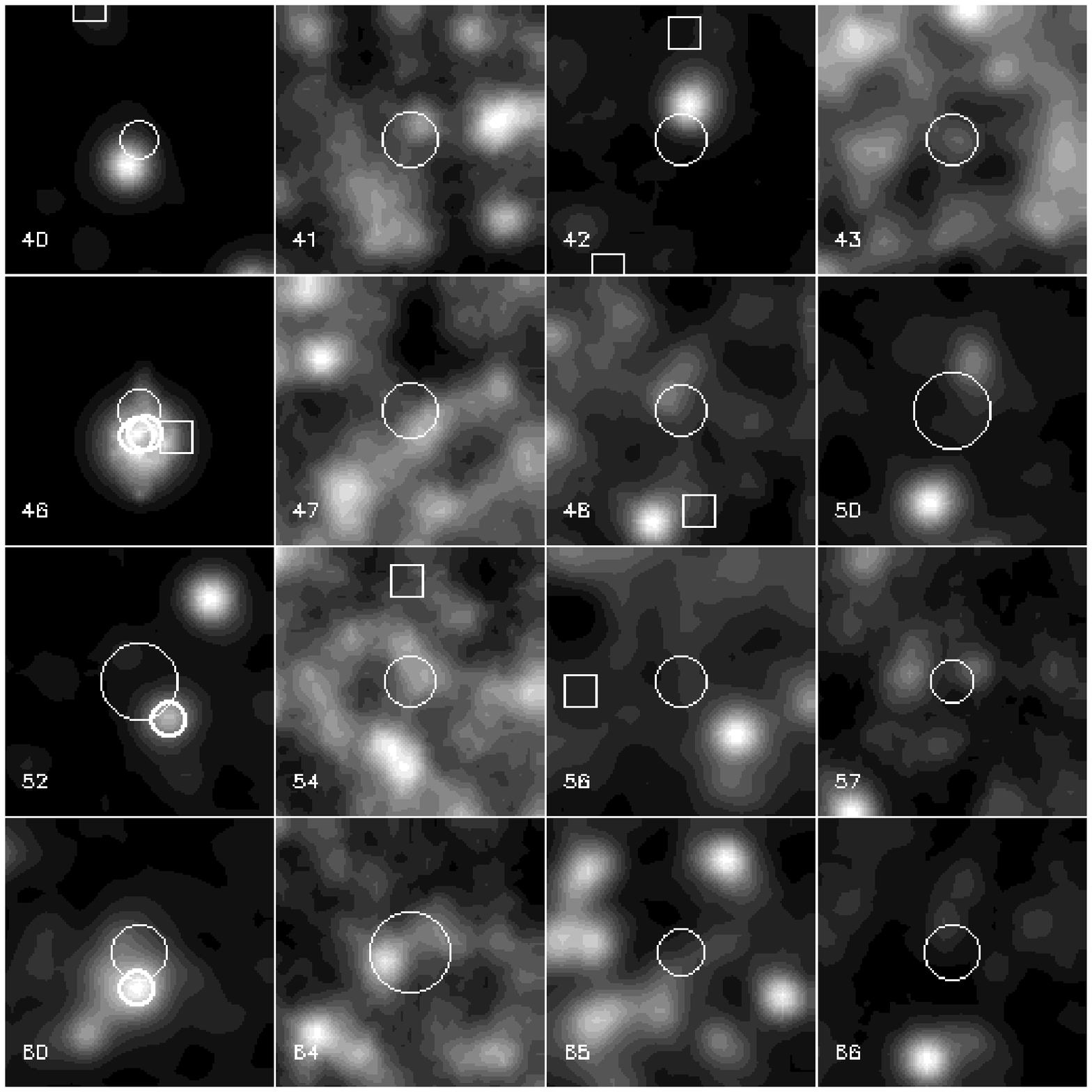}}
 \caption{$7.4''\times 7.4''$ optical images of the region
   centered on the corrected position of the 32 brightest X-ray
   sources inside the $D_{25}$ optical disk.  The circle indicates the
   68\% positional error of the X-ray coordinates, the thicker circles
   show the optical sources found within the X-ray error circle.
   Catalog sources having a distance of less than $5''$ from the X-ray
   position are shown with a box.}
 \label{fig:optic}
\end{figure*}

For the 32 brightest X-ray sources within the $D_{25}$
  ellipse, Fig.~\ref{fig:optic} shows the resulting optical
counterparts in the merged optical image.  As further described in
Sect.~\ref{sec:table} below, we also compare these X-ray and optical
positions with sources from SIMBAD, the USNO-A2.0 catalog, and with
radio sources from \citet{payne:04a}.  We consider sources from these
catalogs as possible counterparts if they have a distance less than
$20''$ from the corrected X-ray positions for X-ray sources, $10''$
for (suspected) supernova remnants and $5''$ for the other sources.
The closest sources of these counterparts are shown in
Fig.~\ref{fig:optic} with a box.

\section{Summary table}
\label{sec:table}

Table~\ref{tab:table} summarizes all information collected from the 86
X-ray sources detected inside the $D_{25}$ disk. The first column
gives the source ID. The second and third columns give the equatorial
sky coordinates of the X-ray sources from the SAS
\texttt{edetect\_chain} task corrected by the \texttt{eposcorr} task.
The combined positional error (in arcsec) from \texttt{edetect\_chain}
and \texttt{eposcorr} is given in column 4.  Column 5 lists the
detection likelihood and columns 6 and 7 give net counts and count
rates, respectively (in counts\,$\text{s}^{-1}$), and their
corresponding uncertainties.  Columns 8 and 9 list the softer and
harder hardness ratios defined by Eq.~\eqref{equ:hard} and their
errors. Column 10 and 11 give the 0.3--6.0\,\kev\ flux and luminosity,
expressed in $\text{erg}\,\text{cm}^{-2}\,\text{s}^{-1}$ and
$\text{erg}\,\text{s}^{-1}$ respectively.

Table~\ref{tab:table2} summarizes all possible optical counterparts
found from the corrected X-ray positions within the X-ray position
error circle.  Column 1 gives the X-ray source ID and the number in
brackets designates a label number when several optical counterparts
are found within the X-ray position error area.  Columns~2 and~3 give
the equatorial sky coordinates of optical counterparts found by
\texttt{idlphot}'s \texttt{find} procedure.  Columns~4, 5, and~6 give
the magnitudes for the optical counterparts, in the B, V, and R band
respectively, with errors of $\sim$0.15\,mag. Column~7 gives
the name and, when available, the reference (within brackets) for
possible radio and optical counterparts sources from SIMBAD (which
includes the \textsl{ROSAT} sources), the USNO-A2.0, and the following
catalogues: \citet[][PFP2004]{payne:04a}, \citet[][SCA2003]{schild},
\citet[][RP2001]{read2001}, \citet[][PGF2001]{piet2001},
\citet[][PDL2000]{Pannuti}, \citet[][BL97]{blair},
\citet[][SMJ96]{soffner}, \citet[][ICS96]{iovino},
\citet[][DCL88]{deharveng}, \citet[][HG86]{humphreys},
\citet[][G84]{graham}.

According to SIMBAD, 14 of our X-ray sources detected inside the
optical disk had already been observed in the X-rays
(labelled `X'), there are 9 SNR or suspected ones (labelled `SNR?'),
11 radio sources (labelled `radio'), from which three are associated
with SNRs and 8 are possible AGNs.  Other sources match with
association of stars (labelled `Assoc*'), H~\textsc{ii} (ionized)
regions (labelled `H~\textsc{ii}'), with regions close to Cepheid
variable stars (labelled `Cepheid'), or with stars (labelled `Star').
Many sources also have an USNO-A2.0 optical counterpart (labelled with
a number).

As already discussed in Sect.~\ref{sec:hard}, our brightest source
(\#1), which has a luminosity of $1.70
\times10^{38}\,\text{erg}\,\text{s}^{-1}$, has been identified by
(source P42) \cite{read2001}, as a possible accreting binary. This
source has been found to have a Wolf Rayet star as optical counterpart
(labelled `WR*'). The source has a luminosity close to
the Eddington limit for a $1.4\,M_\odot$ compact object, which may
suggest the presence of a black hole or neutron star X-ray binary.
Source number 8 has already been identified by \cite{read2001} and
\citet{Kong} as a luminous supersoft X-ray source and has no optical
counterpart.

\setlongtables

\section{Conclusions}
\label{sec:conc}
In this paper we have described the global properties of the detected
X-ray sources in NGC~300 as found in the \textsl{XMM-Newton} data. A
total of 163 sources were found using the 0.3--6\,\kev\ EPIC data of
orbits 192 and 195, of which 86 are located within the optical
$D_{25}$ disk.  This increases the X-ray inventory of NGC~300 by a
factor of $\sim$3.  This increment is mainly due to the better
sensitivity of our observations, caused by the higher
effective area of \textsl{XMM-Newton} with respect to \textsl{ROSAT}.

Using the color-color diagram allowed us to determine the shape of the
X-ray spectrum for each source individually and to estimate the source
fluxes.  The luminosity function of NGC~300 is similar to that of
other spirals \citep{colbert:04a} and can be described by a power law
with slope $1.17\pm 0.17$. It is dominated by soft sources at high
luminosities, although we do not find strong super-Eddington sources
in the galaxy. More information about the brightest X-ray sources
inside the optical disk, such as spectral fitting and
temporal analyses, will be given in a subsequent paper.  The spectrum
of the central diffuse emission region can be described
($\chi^2_\text{red}= 1.52$) by thermal emission from a
  collisionally ionized plasma with $kT=0.2_{-0.01}^{+0.01}$\,\kev,
  plus a second thermal component with a temperature of
  $kT=0.8_{-0.1}^{+0.1}$\,\kev.

The SAS \texttt{eposcorr} task revealed a small positional offset.
After having corrected for this offset, we searched for optical
counterparts in the B, V, and R data and cross-correlate with sources
from SIMBAD and USNO-A2.0 catalogs, and radio sources.

We identified possible optical and radio counterparts to all X-ray
sources using a variety of catalogues.  The brightest X-ray source is
probably a black hole or neutron star X-ray binary, possibly
accreting from a Wolf Rayet star which was identified as the most
likely optical counterpart.  We confirm the presence of a luminous
supersoft X-ray source which has no optical counterpart.

\begin{acknowledgements}
  We thank Jeffrey Payne for kindly providing his list of radio source
  coordinates and for his collaboration. We also thank Wolfgang
  Pietsch for useful discussions and the referee, Andy Read, for his
  thorough refereeing which greatly improved the quality of this
  publication.
  
  This paper is based on observations with \textsl{XMM-Newton}, an ESA
  science mission with instruments and contributions directly financed
  by the ESA Member States and the USA (NASA), and on observations
  made with ESO Telescope at La Silla observatory and retrieved from
  the ESO archive.  We acknowledge partial support from DLR grant
  50OX0002.  This work was furthermore supported by the BMBF through
  the DLR under the project 50OR0106, by the BMBF through DESY under
  the project 05AE2PDA/8, and by the Deutsche Forschungsgemeinschaft
  under the project SCHN 342/3-1.  The support given by ASTROVIRTEL, a
  project funded by the European Commission under FP5 Contract No.\ 
  HPRI-CT-1999-00081, is acknowledged. This research has made use of
  the SIMBAD database, operated at CDS, Strasbourg, France.
\end{acknowledgements}

\onecolumn
\begin{landscape}\scriptsize
\begin{longtable}{lllllllllllllll}
\caption{Summary of the X-ray properties of sources found in NGC~300  (see text for details).\protect\label{tab:table}}\\
\hline 
ID & $\alpha_\text{J2000.0}$ & $\delta_\text{J2000.0}$ & Pos. Err. ($''$) & Lik.& Counts & Ct. rate  & Hard HR & Soft HR &  $F_{0.3-6}$ (\,cgs)  & $L_{0.3-6}$ (\,cgs) \\
(1) & (2) & (3) & (4) & (5) & (6) & (7) & (8) & (9) & (10) & (11) \\
\\ 
\hline
\\
\endfirsthead
\multicolumn{14}{l}{\slshape Continued}   \\
\hline
\\
ID & $\alpha_\text{J2000.0}$ & $\delta_\text{J2000.0}$ & Pos. Err.($''$) & Lik. & Counts & Ct. rate  & Hard HR  & Soft HR &  $F_{0.3-6}$ (\,cgs) &  $L_{0.3-6}$ (\,cgs) \\
(1) & (2) & (3) & (4) & (5) & (6) & (7) & (8) & (9) & (10) & (11) \\

\\
\hline
\\
\endhead

  1 & $ 00:55:10.00 $ & $ -37:42:12.0 $ & $   0.45 $ & $ 5.45\times 10^{4} $ & $   1.53 \pm   0.01\times 10^{4} $ & $   7.39 \pm   0.06\times 10^{-2} $ & $  -0.18  \pm        0.01 $ & $  -0.29  \pm        0.01 $ & $   3.49_{-  0.12}^{+  0.03}\times 10^{-13} $ & $   1.70\times 10^{38}$ \\ [3pt]
  3 & $ 00:55:10.86 $ & $ -37:48:34.4 $ & $   0.47 $ & $ 8.30\times 10^{3} $ & $   3.16 \pm   0.06\times 10^{3} $ & $   1.61 \pm   0.03\times 10^{-2} $ & $  -0.21  \pm        0.01 $ & $  -0.30  \pm        0.01 $ & $   7.20_{-  0.74}^{+  0.33}\times 10^{-14} $ & $   3.51\times 10^{37}$ \\ [3pt]
  7 & $ 00:54:50.33 $ & $ -37:38:49.5 $ & $   0.47 $ & $ 7.43\times 10^{3} $ & $   2.94 \pm   0.06\times 10^{3} $ & $   1.50 \pm   0.03\times 10^{-2} $ & $  -0.14  \pm        0.02 $ & $ \phantom{+}  0.28  \pm        0.02 $ & $   1.06_{-  0.05}^{+  0.04}\times 10^{-13} $ & $   5.20\times 10^{37}$ \\ [3pt]
  8 & $ 00:55:10.91 $ & $ -37:38:54.3 $ & $   0.46 $ & $ 3.85\times 10^{3} $ & $   1.90 \pm   0.15\times 10^{2} $ & $   4.31 \pm   0.34\times 10^{-3} $ & $ \phantom{+}  0.01  \pm        0.03 $ & $  -0.97  \pm        0.03 $ & $   1.52_{-  1.30}^{+  0.56}\times 10^{-14} $ & $   7.40\times 10^{36}$ \\ [3pt]
 12 & $ 00:54:40.66 $ & $ -37:40:48.9 $ & $   0.58 $ & $ 1.20\times 10^{3} $ & $   3.85 \pm   0.22\times 10^{2} $ & $   2.83 \pm   0.16\times 10^{-3} $ & $  -0.17  \pm        0.03 $ & $  -0.58  \pm        0.03 $ & $   1.03_{-  0.09}^{+  0.12}\times 10^{-14} $ & $   5.03\times 10^{36}$ \\ [3pt]
 13 & $ 00:55:27.50 $ & $ -37:36:53.5 $ & $   0.56 $ & $ 9.50\times 10^{2} $ & $   5.38 \pm   0.27\times 10^{2} $ & $   2.72 \pm   0.14\times 10^{-3} $ & $  -0.12  \pm        0.04 $ & $ \phantom{+}  0.11  \pm        0.04 $ & $   1.82_{-  0.19}^{+  0.18}\times 10^{-14} $ & $   8.89\times 10^{36}$ \\ [3pt]
 17 & $ 00:54:45.22 $ & $ -37:41:46.7 $ & $   0.53 $ & $ 1.45\times 10^{3} $ & $   2.73 \pm   0.19\times 10^{2} $ & $   5.26 \pm   0.35\times 10^{-3} $ & $  -0.15  \pm        0.03 $ & $  -0.67  \pm        0.03 $ & $   1.91_{-  0.32}^{+  0.26}\times 10^{-14} $ & $   9.32\times 10^{36}$ \\ [3pt]
 20 & $ 00:55:20.47 $ & $ -37:48:10.2 $ & $   0.69 $ & $ 3.75\times 10^{2} $ & $   2.75 \pm   0.19\times 10^{2} $ & $   1.60 \pm   0.11\times 10^{-3} $ & $  -0.23  \pm        0.04 $ & $  -0.41  \pm        0.04 $ & $   6.36_{-  1.03}^{+  0.98}\times 10^{-15} $ & $   3.11\times 10^{36}$ \\ [3pt]
 21 & $ 00:54:33.96 $ & $ -37:44:43.2 $ & $   0.58 $ & $ 9.17\times 10^{2} $ & $   5.65 \pm   0.27\times 10^{2} $ & $   2.82 \pm   0.14\times 10^{-3} $ & $  -0.13  \pm        0.04 $ & $  -0.01  \pm        0.04 $ & $   1.70_{-  0.11}^{+  0.23}\times 10^{-14} $ & $   8.32\times 10^{36}$ \\ [3pt]
 26 & $ 00:55:26.23 $ & $ -37:38:38.2 $ & $   0.69 $ & $ 4.01\times 10^{2} $ & $   3.01 \pm   0.20\times 10^{2} $ & $   1.57 \pm   0.10\times 10^{-3} $ & $  -0.08  \pm        0.05 $ & $ \phantom{+}  0.04  \pm        0.05 $ & $   1.08_{-  0.15}^{+  0.12}\times 10^{-14} $ & $   5.29\times 10^{36}$ \\ [3pt]
 28 & $ 00:54:42.56 $ & $ -37:43:43.2 $ & $   0.61 $ & $ 6.38\times 10^{2} $ & $   3.98 \pm   0.24\times 10^{2} $ & $   1.98 \pm   0.12\times 10^{-3} $ & $  -0.13  \pm        0.05 $ & $ \phantom{+}  0.06  \pm        0.05 $ & $   1.28_{-  0.16}^{+  0.18}\times 10^{-14} $ & $   6.23\times 10^{36}$ \\ [3pt]
 31 & $ 00:54:30.57 $ & $ -37:43:15.9 $ & $   0.65 $ & $ 5.11\times 10^{2} $ & $   3.26 \pm   0.21\times 10^{2} $ & $   1.67 \pm   0.11\times 10^{-3} $ & $  -0.07  \pm        0.05 $ & $ \phantom{+}  0.00  \pm        0.05 $ & $   1.14_{-  0.15}^{+  0.13}\times 10^{-14} $ & $   5.58\times 10^{36}$ \\ [3pt]
 34 & $ 00:55:15.39 $ & $ -37:44:38.9 $ & $   0.70 $ & $ 3.78\times 10^{2} $ & $   6.98 \pm   0.97\times 10^{1} $ & $   9.66 \pm   1.29\times 10^{-4} $ & $  -0.11  \pm        0.08 $ & $  -0.60  \pm        0.08 $ & $   3.96_{-  0.95}^{+  1.97}\times 10^{-15} $ & $   1.93\times 10^{36}$ \\ [3pt]
 37 & $ 00:55:17.52 $ & $ -37:44:55.8 $ & $   0.66 $ & $ 3.82\times 10^{2} $ & $   2.65 \pm   0.19\times 10^{2} $ & $   1.37 \pm   0.10\times 10^{-3} $ & $  -0.27  \pm        0.06 $ & $ \phantom{+}  0.09  \pm        0.06 $ & $   7.41_{-  1.21}^{+  1.03}\times 10^{-15} $ & $   3.62\times 10^{36}$ \\ [3pt]
 38 & $ 00:54:42.59 $ & $ -37:37:32.7 $ & $   0.69 $ & $ 3.77\times 10^{2} $ & $   2.75 \pm   0.20\times 10^{2} $ & $   1.38 \pm   0.10\times 10^{-3} $ & $  -0.21  \pm        0.05 $ & $  -0.11  \pm        0.05 $ & $   6.97_{-  0.63}^{+  1.45}\times 10^{-15} $ & $   3.40\times 10^{36}$ \\ [3pt]
 39 & $ 00:54:49.55 $ & $ -37:40:00.7 $ & $   0.68 $ & $ 3.28\times 10^{2} $ & $   2.85 \pm   0.20\times 10^{2} $ & $   1.37 \pm   0.10\times 10^{-3} $ & $  -0.12  \pm        0.06 $ & $ \phantom{+}  0.11  \pm        0.06 $ & $   9.21_{-  1.35}^{+  1.51}\times 10^{-15} $ & $   4.50\times 10^{36}$ \\ [3pt]
 40 & $ 00:54:13.89 $ & $ -37:37:10.8 $ & $   0.76 $ & $ 2.17\times 10^{2} $ & $   1.97 \pm   0.17\times 10^{2} $ & $   9.65 \pm   0.83\times 10^{-4} $ & $  -0.18  \pm        0.07 $ & $ \phantom{+}  0.15  \pm        0.07 $ & $   6.21_{-  1.16}^{+  1.14}\times 10^{-15} $ & $   3.03\times 10^{36}$ \\ [3pt]
 41 & $ 00:54:48.07 $ & $ -37:46:57.4 $ & $   0.79 $ & $ 3.27\times 10^{2} $ & $   2.41 \pm   0.19\times 10^{2} $ & $   1.18 \pm   0.09\times 10^{-3} $ & $  -0.09  \pm        0.06 $ & $  -0.01  \pm        0.06 $ & $   7.71_{-  1.16}^{+  1.35}\times 10^{-15} $ & $   3.77\times 10^{36}$ \\ [3pt]
 42 & $ 00:54:48.11 $ & $ -37:45:40.0 $ & $   0.69 $ & $ 3.71\times 10^{2} $ & $   2.62 \pm   0.19\times 10^{2} $ & $   1.26 \pm   0.09\times 10^{-3} $ & $  -0.09  \pm        0.06 $ & $ \phantom{+}  0.12  \pm        0.06 $ & $   8.82_{-  1.27}^{+  1.46}\times 10^{-15} $ & $   4.30\times 10^{36}$ \\ [3pt]
 43 & $ 00:54:57.28 $ & $ -37:43:11.4 $ & $   0.69 $ & $ 3.38\times 10^{2} $ & $   2.81 \pm   0.20\times 10^{2} $ & $   1.41 \pm   0.10\times 10^{-3} $ & $  -0.09  \pm        0.06 $ & $ \phantom{+}  0.06  \pm        0.06 $ & $   9.75_{-  1.64}^{+  1.15}\times 10^{-15} $ & $   4.76\times 10^{36}$ \\ [3pt]
 46 & $ 00:55:42.91 $ & $ -37:44:34.8 $ & $   0.96 $ & $ 1.33\times 10^{2} $ & $   6.77 \pm   1.05\times 10^{1} $ & $   5.89 \pm   0.88\times 10^{-4} $ & $  -0.01  \pm        0.09 $ & $  -0.61  \pm        0.09 $ & $   3.29_{-  0.99}^{+  1.00}\times 10^{-15} $ & $   1.61\times 10^{36}$ \\ [3pt]
 47 & $ 00:54:50.63 $ & $ -37:46:00.7 $ & $   0.69 $ & $ 2.51\times 10^{2} $ & $   1.99 \pm   0.17\times 10^{2} $ & $   9.48 \pm   0.80\times 10^{-4} $ & $  -0.02  \pm        0.07 $ & $ \phantom{+}  0.09  \pm        0.07 $ & $   7.18_{-  1.13}^{+  1.25}\times 10^{-15} $ & $   3.51\times 10^{36}$ \\ [3pt]
 48 & $ 00:54:19.97 $ & $ -37:39:08.9 $ & $   0.87 $ & $ 1.63\times 10^{2} $ & $   1.61 \pm   0.16\times 10^{2} $ & $   7.16 \pm   0.73\times 10^{-4} $ & $  -0.12  \pm        0.09 $ & $ \phantom{+}  0.32  \pm        0.09 $ & $   5.36_{-  1.03}^{+  1.17}\times 10^{-15} $ & $   2.62\times 10^{36}$ \\ [3pt]
 50 & $ 00:55:07.67 $ & $ -37:44:20.1 $ & $   0.76 $ & $ 2.13\times 10^{2} $ & $   2.01 \pm   0.17\times 10^{2} $ & $   1.01 \pm   0.09\times 10^{-3} $ & $ \phantom{+}  0.01  \pm        0.08 $ & $ \phantom{+}  0.20  \pm        0.08 $ & $   8.34_{-  1.31}^{+  1.48}\times 10^{-15} $ & $   4.07\times 10^{36}$ \\ [3pt]
 52 & $ 00:55:42.24 $ & $ -37:40:23.7 $ & $   1.13 $ & $ 9.98\times 10^{1} $ & $   6.52 \pm   1.03\times 10^{1} $ & $   1.14 \pm   0.17\times 10^{-3} $ & $  -0.01  \pm        0.07 $ & $  -0.86  \pm        0.07 $ & $   5.08_{-  1.38}^{+  2.02}\times 10^{-15} $ & $   2.48\times 10^{36}$ \\ [3pt]
 54 & $ 00:54:41.04 $ & $ -37:33:51.7 $ & $   0.99 $ & $ 1.39\times 10^{2} $ & $   1.46 \pm   0.14\times 10^{2} $ & $   7.92 \pm   0.81\times 10^{-4} $ & $ \phantom{+}  0.35  \pm        0.10 $ & $ \phantom{+}  0.12  \pm        0.10 $ & $   8.98_{-  1.49}^{+  1.88}\times 10^{-15} $ & $   4.38\times 10^{36}$ \\ [3pt]
 56 & $ 00:54:50.56 $ & $ -37:41:27.8 $ & $   0.78 $ & $ 2.16\times 10^{2} $ & $   1.37 \pm   0.14\times 10^{2} $ & $   8.59 \pm   0.91\times 10^{-4} $ & $  -0.16  \pm        0.07 $ & $  -0.44  \pm        0.07 $ & $   3.62_{-  0.67}^{+  1.04}\times 10^{-15} $ & $   1.77\times 10^{36}$ \\ [3pt]
 57 & $ 00:55:27.24 $ & $ -37:36:13.5 $ & $   1.06 $ & $ 9.43\times 10^{1} $ & $   1.10 \pm   0.13\times 10^{2} $ & $   6.76 \pm   0.79\times 10^{-4} $ & $  -0.17  \pm        0.09 $ & $ \phantom{+}  0.06  \pm        0.09 $ & $   4.19_{-  0.95}^{+  1.03}\times 10^{-15} $ & $   2.05\times 10^{36}$ \\ [3pt]
 60 & $ 00:54:44.38 $ & $ -37:41:14.7 $ & $   0.78 $ & $ 1.65\times 10^{2} $ & $   8.66 \pm   1.15\times 10^{1} $ & $   1.20 \pm   0.16\times 10^{-3} $ & $  -0.16  \pm        0.07 $ & $  -0.51  \pm        0.07 $ & $   5.01_{-  1.30}^{+  1.58}\times 10^{-15} $ & $   2.45\times 10^{36}$ \\ [3pt]
 64 & $ 00:55:27.16 $ & $ -37:45:16.5 $ & $   0.95 $ & $ 1.04\times 10^{2} $ & $   1.02 \pm   0.13\times 10^{2} $ & $   6.52 \pm   0.78\times 10^{-4} $ & $  -0.33  \pm        0.10 $ & $ \phantom{+}  0.08  \pm        0.10 $ & $   3.19_{-  0.76}^{+  1.02}\times 10^{-15} $ & $   1.56\times 10^{36}$ \\ [3pt]
 65 & $ 00:54:51.52 $ & $ -37:35:34.5 $ & $   0.98 $ & $ 1.14\times 10^{2} $ & $   1.01 \pm   0.12\times 10^{2} $ & $   7.63 \pm   0.94\times 10^{-4} $ & $  -0.07  \pm        0.10 $ & $ \phantom{+}  0.02  \pm        0.10 $ & $   5.22_{-  1.18}^{+  1.46}\times 10^{-15} $ & $   2.55\times 10^{36}$ \\ [3pt]
 66 & $ 00:54:33.79 $ & $ -37:44:26.1 $ & $   0.94 $ & $ 8.00\times 10^{1} $ & $   2.01 \pm   0.17\times 10^{2} $ & $   9.44 \pm   0.84\times 10^{-4} $ & $  -0.07  \pm        0.08 $ & $ \phantom{+}  0.35  \pm        0.08 $ & $   7.49_{-  1.32}^{+  1.45}\times 10^{-15} $ & $   3.66\times 10^{36}$ \\ [3pt]
 67 & $ 00:54:35.93 $ & $ -37:34:33.8 $ & $   1.02 $ & $ 8.91\times 10^{1} $ & $   1.02 \pm   0.13\times 10^{2} $ & $   5.10 \pm   0.64\times 10^{-4} $ & $  -0.03  \pm        0.10 $ & $ \phantom{+}  0.05  \pm        0.10 $ & $   3.68_{-  0.83}^{+  0.99}\times 10^{-15} $ & $   1.80\times 10^{36}$ \\ [3pt]
 69 & $ 00:54:31.67 $ & $ -37:38:27.7 $ & $   0.98 $ & $ 9.87\times 10^{1} $ & $   9.22 \pm   1.31\times 10^{1} $ & $   4.55 \pm   0.64\times 10^{-4} $ & $  -0.17  \pm        0.11 $ & $ \phantom{+}  0.12  \pm        0.11 $ & $   2.88_{-  0.77}^{+  0.90}\times 10^{-15} $ & $   1.40\times 10^{36}$ \\ [3pt]
 71 & $ 00:55:11.28 $ & $ -37:46:37.7 $ & $   0.97 $ & $ 1.01\times 10^{2} $ & $   9.27 \pm   1.24\times 10^{1} $ & $   5.75 \pm   0.74\times 10^{-4} $ & $  -0.18  \pm        0.12 $ & $ \phantom{+}  0.43  \pm        0.12 $ & $   4.18_{-  1.08}^{+  1.08}\times 10^{-15} $ & $   2.04\times 10^{36}$ \\ [3pt]
 72 & $ 00:54:37.68 $ & $ -37:42:49.6 $ & $   0.92 $ & $ 1.29\times 10^{2} $ & $   1.34 \pm   0.14\times 10^{2} $ & $   6.91 \pm   0.73\times 10^{-4} $ & $  -0.09  \pm        0.08 $ & $  -0.01  \pm        0.08 $ & $   4.53_{-  0.86}^{+  1.17}\times 10^{-15} $ & $   2.21\times 10^{36}$ \\ [3pt]
 73 & $ 00:55:31.08 $ & $ -37:37:56.1 $ & $   1.20 $ & $ 5.32\times 10^{1} $ & $   6.68 \pm   1.01\times 10^{1} $ & $   4.32 \pm   0.63\times 10^{-4} $ & $  -0.05  \pm        0.12 $ & $ \phantom{+}  0.10  \pm        0.12 $ & $   3.14_{-  0.87}^{+  1.10}\times 10^{-15} $ & $   1.53\times 10^{36}$ \\ [3pt]
 74 & $ 00:54:47.71 $ & $ -37:32:57.5 $ & $   1.45 $ & $ 4.19\times 10^{1} $ & $   4.24 \pm   0.87\times 10^{1} $ & $   3.33 \pm   0.63\times 10^{-4} $ & $ \phantom{+}  0.05  \pm        0.15 $ & $  -0.06  \pm        0.15 $ & $   2.56_{-  0.86}^{+  1.04}\times 10^{-15} $ & $   1.25\times 10^{36}$ \\ [3pt]
 79 & $ 00:54:22.13 $ & $ -37:40:25.0 $ & $   1.05 $ & $ 9.44\times 10^{1} $ & $   4.04 \pm   3.11\times 10^{0} $ & $   1.79 \pm   1.03\times 10^{-4} $ & $    $ & $    $ & $     $ & $  $ \\ [3pt]
 84 & $ 00:54:25.03 $ & $ -37:43:56.3 $ & $   1.04 $ & $ 8.13\times 10^{1} $ & $   5.38 \pm   0.93\times 10^{1} $ & $   3.69 \pm   0.72\times 10^{-4} $ & $ \phantom{+}  0.27  \pm        0.19 $ & $ \phantom{+}  0.31  \pm        0.19 $ & $   3.99_{-  1.25}^{+  1.37}\times 10^{-15} $ & $   1.95\times 10^{36}$ \\ [3pt]
 87 & $ 00:54:12.03 $ & $ -37:39:51.7 $ & $   1.08 $ & $ 6.42\times 10^{1} $ & $   8.52 \pm   1.14\times 10^{1} $ & $   4.73 \pm   0.65\times 10^{-4} $ & $  -0.19  \pm        0.12 $ & $ \phantom{+}  0.20  \pm        0.12 $ & $   3.09_{-  0.88}^{+  1.01}\times 10^{-15} $ & $   1.51\times 10^{36}$ \\ [3pt]
 88 & $ 00:54:22.44 $ & $ -37:43:13.3 $ & $   1.02 $ & $ 9.19\times 10^{1} $ & $   3.74 \pm   0.81\times 10^{1} $ & $   6.33 \pm   1.18\times 10^{-4} $ & $  -0.08  \pm        0.12 $ & $  -0.41  \pm        0.12 $ & $   3.24_{-  1.19}^{+  1.76}\times 10^{-15} $ & $   1.58\times 10^{36}$ \\ [3pt]
 90 & $ 00:54:53.28 $ & $ -37:41:27.0 $ & $   0.95 $ & $ 8.94\times 10^{1} $ & $   9.37 \pm   1.27\times 10^{1} $ & $   5.80 \pm   0.78\times 10^{-4} $ & $  -0.16  \pm        0.10 $ & $  -0.06  \pm        0.10 $ & $   3.37_{-  0.86}^{+  1.06}\times 10^{-15} $ & $   1.65\times 10^{36}$ \\ [3pt]
 91 & $ 00:54:25.32 $ & $ -37:44:39.5 $ & $   1.11 $ & $ 8.24\times 10^{1} $ & $   8.25 \pm   1.17\times 10^{1} $ & $   6.29 \pm   0.87\times 10^{-4} $ & $  -0.14  \pm        0.11 $ & $  -0.07  \pm        0.11 $ & $   3.66_{-  0.95}^{+  1.16}\times 10^{-15} $ & $   1.79\times 10^{36}$ \\ [3pt]
 92 & $ 00:54:21.04 $ & $ -37:42:40.8 $ & $   1.12 $ & $ 6.19\times 10^{1} $ & $   4.58 \pm   0.83\times 10^{1} $ & $   3.89 \pm   0.76\times 10^{-4} $ & $ \phantom{+}  0.23  \pm        0.18 $ & $ \phantom{+}  0.18  \pm        0.18 $ & $   4.04_{-  1.36}^{+  1.63}\times 10^{-15} $ & $   1.97\times 10^{36}$ \\ [3pt]
 94 & $ 00:55:12.23 $ & $ -37:38:23.8 $ & $   1.10 $ & $ 5.98\times 10^{1} $ & $   7.99 \pm   1.16\times 10^{1} $ & $   5.85 \pm   0.88\times 10^{-4} $ & $  -0.09  \pm        0.10 $ & $  -0.33  \pm        0.10 $ & $   3.07_{-  0.90}^{+  1.12}\times 10^{-15} $ & $   1.50\times 10^{36}$ \\ [3pt]
 99 & $ 00:54:47.35 $ & $ -37:48:26.9 $ & $   1.13 $ & $ 5.72\times 10^{1} $ & $   5.39 \pm   0.97\times 10^{1} $ & $   4.23 \pm   0.73\times 10^{-4} $ & $  -0.05  \pm        0.15 $ & $ \phantom{+}  0.17  \pm        0.15 $ & $   3.21_{-  1.02}^{+  1.23}\times 10^{-15} $ & $   1.57\times 10^{36}$ \\ [3pt]
100 & $ 00:54:12.47 $ & $ -37:43:20.7 $ & $   1.23 $ & $ 4.86\times 10^{1} $ & $   6.01 \pm   1.03\times 10^{1} $ & $   3.26 \pm   0.53\times 10^{-4} $ & $  -0.31  \pm        0.14 $ & $ \phantom{+}  0.19  \pm        0.14 $ & $   1.77_{-  0.59}^{+  0.70}\times 10^{-15} $ & $   8.64\times 10^{35}$ \\ [3pt]
102 & $ 00:54:28.55 $ & $ -37:41:29.7 $ & $   1.04 $ & $ 6.83\times 10^{1} $ & $   1.09 \pm   0.39\times 10^{1} $ & $   3.53 \pm   1.27\times 10^{-4} $ & $    $ & $    $ & $     $ & $  $ \\ [3pt]
103 & $ 00:55:03.66 $ & $ -37:37:40.1 $ & $   1.00 $ & $ 6.14\times 10^{1} $ & $   8.42 \pm   1.11\times 10^{1} $ & $   4.08 \pm   0.57\times 10^{-4} $ & $ \phantom{+}  0.00  \pm        0.09 $ & $  -0.42  \pm        0.09 $ & $   2.36_{-  0.62}^{+  0.88}\times 10^{-15} $ & $   1.15\times 10^{36}$ \\ [3pt]
107 & $ 00:54:59.02 $ & $ -37:47:51.0 $ & $   1.42 $ & $ 4.69\times 10^{1} $ & $   5.35 \pm   0.96\times 10^{1} $ & $   4.01 \pm   0.70\times 10^{-4} $ & $ \phantom{+}  0.34  \pm        0.17 $ & $ \phantom{+}  0.24  \pm        0.17 $ & $   4.58_{-  1.31}^{+  1.46}\times 10^{-15} $ & $   2.23\times 10^{36}$ \\ [3pt]
112 & $ 00:54:15.91 $ & $ -37:45:07.0 $ & $   1.42 $ & $ 4.45\times 10^{1} $ & $   4.59 \pm   0.86\times 10^{1} $ & $   4.61 \pm   0.86\times 10^{-4} $ & $  -0.03  \pm        0.09 $ & $  -0.74  \pm        0.09 $ & $   2.26_{-  0.99}^{+  0.71}\times 10^{-15} $ & $   1.10\times 10^{36}$ \\ [3pt]
117 & $ 00:55:13.85 $ & $ -37:47:57.8 $ & $   1.71 $ & $ 2.49\times 10^{1} $ & $   3.39 \pm   0.77\times 10^{1} $ & $   3.04 \pm   0.71\times 10^{-4} $ & $  -0.08  \pm        0.19 $ & $ \phantom{+}  0.19  \pm        0.19 $ & $   2.23_{-  0.86}^{+  1.26}\times 10^{-15} $ & $   1.09\times 10^{36}$ \\ [3pt]
120 & $ 00:54:53.26 $ & $ -37:43:10.8 $ & $   1.07 $ & $ 5.21\times 10^{1} $ & $   4.63 \pm   0.85\times 10^{1} $ & $   3.85 \pm   0.74\times 10^{-4} $ & $ \phantom{+}  0.40  \pm        0.17 $ & $  -0.31  \pm        0.17 $ &  & \\ [3pt]
122 & $ 00:55:29.75 $ & $ -37:37:27.4 $ & $   1.65 $ & $ 1.80\times 10^{1} $ & $   1.14 \pm   0.49\times 10^{1} $ & $   1.58 \pm   0.64\times 10^{-4} $ & $    $ & $    $ & $     $ & $  $ \\ [3pt]
123 & $ 00:55:06.00 $ & $ -37:41:18.1 $ & $   1.62 $ & $ 2.37\times 10^{1} $ & $   5.79 \pm   3.03\times 10^{0} $ & $   1.74 \pm   0.87\times 10^{-4} $ & $    $ & $    $ & $     $ & $  $ \\ [3pt]
125 & $ 00:55:34.06 $ & $ -37:46:35.1 $ & $   1.66 $ & $ 1.81\times 10^{1} $ & $   2.61 \pm   0.65\times 10^{1} $ & $   2.26 \pm   0.62\times 10^{-4} $ & $  -0.21  \pm        0.24 $ & $ \phantom{+}  0.32  \pm        0.24 $ & $   1.51_{-  0.69}^{+  1.05}\times 10^{-15} $ & $   7.40\times 10^{35}$ \\ [3pt]
126 & $ 00:54:53.47 $ & $ -37:40:28.3 $ & $   1.42 $ & $ 3.10\times 10^{1} $ & $   2.46 \pm   0.65\times 10^{1} $ & $   2.27 \pm   0.59\times 10^{-4} $ & $  -0.10  \pm        0.16 $ & $  -0.53  \pm        0.16 $ & $   1.04_{-  0.48}^{+  0.81}\times 10^{-15} $ & $   5.09\times 10^{35}$ \\ [3pt]
128 & $ 00:55:23.99 $ & $ -37:45:24.1 $ & $   1.51 $ & $ 2.41\times 10^{1} $ & $   3.23 \pm   0.74\times 10^{1} $ & $   2.25 \pm   0.54\times 10^{-4} $ & $ \phantom{+}  0.18  \pm        0.21 $ & $ \phantom{+}  0.13  \pm        0.21 $ & $   2.19_{-  0.89}^{+  1.06}\times 10^{-15} $ & $   1.07\times 10^{36}$ \\ [3pt]
132 & $ 00:54:30.56 $ & $ -37:40:05.4 $ & $   1.39 $ & $ 3.09\times 10^{1} $ & $   2.89 \pm   0.72\times 10^{1} $ & $   3.02 \pm   0.96\times 10^{-4} $ & $ \phantom{+}  0.48  \pm        0.31 $ & $ \phantom{+}  0.15  \pm        0.31 $ & $   3.87_{-  1.88}^{+  1.49}\times 10^{-15} $ & $   1.89\times 10^{36}$ \\ [3pt]
134 & $ 00:55:14.88 $ & $ -37:48:51.2 $ & $   1.87 $ & $ 1.25\times 10^{1} $ & $   1.54 \pm   0.56\times 10^{1} $ & $   1.29 \pm   0.49\times 10^{-4} $ & $    $ & $    $ & $     $ & $  $ \\ [3pt]
135 & $ 00:54:31.31 $ & $ -37:45:26.2 $ & $   1.44 $ & $ 2.91\times 10^{1} $ & $   4.52 \pm   2.89\times 10^{0} $ & $   1.85 \pm   0.96\times 10^{-4} $ & $    $ & $    $ & $     $ & $  $ \\ [3pt]
136 & $ 00:54:44.82 $ & $ -37:37:42.1 $ & $   1.38 $ & $ 2.07\times 10^{1} $ & $   2.12 \pm   0.62\times 10^{1} $ & $   1.81 \pm   0.55\times 10^{-4} $ & $  -0.16  \pm        0.24 $ & $ \phantom{+}  0.20  \pm        0.24 $ & $   1.22_{-  0.61}^{+  0.88}\times 10^{-15} $ & $   5.97\times 10^{35}$ \\ [3pt]
137 & $ 00:54:02.59 $ & $ -37:39:31.4 $ & $   1.96 $ & $ 1.05\times 10^{1} $ & $   1.09 \pm   0.42\times 10^{1} $ & $   1.68 \pm   0.66\times 10^{-4} $ & $    $ & $    $ & $     $ & $  $ \\ [3pt]
139 & $ 00:55:07.51 $ & $ -37:45:14.6 $ & $   1.61 $ & $ 2.24\times 10^{1} $ & $   1.97 \pm   0.61\times 10^{1} $ & $   2.15 \pm   0.58\times 10^{-4} $ & $    $ & $    $ & $     $ & $  $ \\ [3pt]
140 & $ 00:55:26.77 $ & $ -37:38:12.7 $ & $   1.84 $ & $ 1.40\times 10^{1} $ & $   1.75 \pm   0.52\times 10^{1} $ & $   1.64 \pm   0.44\times 10^{-4} $ & $    $ & $    $ & $     $ & $  $ \\ [3pt]
141 & $ 00:54:56.97 $ & $ -37:47:26.7 $ & $   1.49 $ & $ 2.47\times 10^{1} $ & $   1.47 \pm   0.50\times 10^{1} $ & $   3.11 \pm   0.94\times 10^{-4} $ & $    $ & $    $ & $     $ & $  $ \\ [3pt]
142 & $ 00:54:33.17 $ & $ -37:44:03.9 $ & $   1.78 $ & $ 1.85\times 10^{1} $ & $   1.15 \pm   0.45\times 10^{1} $ & $   1.33 \pm   0.56\times 10^{-4} $ & $    $ & $    $ & $     $ & $  $ \\ [3pt]
143 & $ 00:55:31.39 $ & $ -37:40:00.1 $ & $   1.87 $ & $ 1.44\times 10^{1} $ & $   1.13 \pm   0.39\times 10^{1} $ & $   1.13 \pm   0.47\times 10^{-4} $ & $    $ & $    $ & $     $ & $  $ \\ [3pt]
145 & $ 00:55:01.11 $ & $ -37:34:39.4 $ & $   2.96 $ & $ 1.19\times 10^{1} $ & $   6.33 \pm   3.98\times 10^{0} $ & $   1.18 \pm   0.66\times 10^{-4} $ & $    $ & $    $ & $     $ & $  $ \\ [3pt]
146 & $ 00:54:57.44 $ & $ -37:45:36.3 $ & $   1.49 $ & $ 2.57\times 10^{1} $ & $   3.43 \pm   0.77\times 10^{1} $ & $   2.33 \pm   0.48\times 10^{-4} $ & $ \phantom{+}  0.00  \pm        0.16 $ & $  -0.01  \pm        0.16 $ & $   1.74_{-  0.64}^{+  0.74}\times 10^{-15} $ & $   8.48\times 10^{35}$ \\ [3pt]
147 & $ 00:54:51.84 $ & $ -37:47:08.2 $ & $   1.76 $ & $ 1.75\times 10^{1} $ & $   2.63 \pm   0.62\times 10^{1} $ & $   1.93 \pm   0.47\times 10^{-4} $ & $  -0.23  \pm        0.19 $ & $  -0.09  \pm        0.19 $ & $   1.01_{-  0.47}^{+  0.61}\times 10^{-15} $ & $   4.93\times 10^{35}$ \\ [3pt]
148 & $ 00:55:03.32 $ & $ -37:45:37.6 $ & $   1.79 $ & $ 1.74\times 10^{1} $ & $   2.43 \pm   0.68\times 10^{1} $ & $   1.40 \pm   0.41\times 10^{-4} $ & $  -0.09  \pm        0.20 $ & $  -0.34  \pm        0.20 $ & $   7.09_{-  3.30}^{+  5.97}\times 10^{-16} $ & $   3.46\times 10^{35}$ \\ [3pt]
150 & $ 00:55:06.32 $ & $ -37:37:53.2 $ & $   1.78 $ & $ 1.51\times 10^{1} $ & $   1.59 \pm   0.57\times 10^{1} $ & $   1.44 \pm   0.56\times 10^{-4} $ & $    $ & $    $ & $     $ & $  $ \\ [3pt]
151 & $ 00:55:33.79 $ & $ -37:43:11.0 $ & $   2.74 $ & $ 1.01\times 10^{1} $ & $   0.00 \pm   0.00\times 10^{0} $ & $    $ & $    $ & $    $ & $     $ & $  $ \\ [3pt]
152 & $ 00:55:05.57 $ & $ -37:42:41.4 $ & $   1.79 $ & $ 1.07\times 10^{1} $ & $   1.43 \pm   0.56\times 10^{1} $ & $   1.68 \pm   0.63\times 10^{-4} $ & $    $ & $    $ & $     $ & $  $ \\ [3pt]
153 & $ 00:54:07.51 $ & $ -37:41:15.6 $ & $   2.10 $ & $ 1.02\times 10^{1} $ & $   0.00 \pm   0.00\times 10^{0} $ & $    $ & $    $ & $    $ & $     $ & $  $ \\ [3pt]
155 & $ 00:54:56.89 $ & $ -37:43:39.3 $ & $   1.60 $ & $ 1.63\times 10^{1} $ & $   2.16 \pm   0.62\times 10^{1} $ & $   1.38 \pm   0.38\times 10^{-4} $ & $  -0.16  \pm        0.22 $ & $  -0.07  \pm        0.22 $ & $   7.73_{-  3.69}^{+  5.59}\times 10^{-16} $ & $   3.77\times 10^{35}$ \\ [3pt]
156 & $ 00:54:29.92 $ & $ -37:40:31.5 $ & $   1.55 $ & $ 1.46\times 10^{1} $ & $   1.21 \pm   0.44\times 10^{1} $ & $   1.49 \pm   0.50\times 10^{-4} $ & $    $ & $    $ & $     $ & $  $ \\ [3pt]
157 & $ 00:54:52.65 $ & $ -37:46:00.7 $ & $   2.23 $ & $ 1.13\times 10^{1} $ & $   0.00 \pm   0.00\times 10^{0} $ & $    $ & $    $ & $    $ & $     $ & $  $ \\ [3pt]
158 & $ 00:54:53.33 $ & $ -37:44:40.1 $ & $   1.74 $ & $ 1.30\times 10^{1} $ & $   1.25 \pm   0.49\times 10^{1} $ & $   8.99 \pm   3.67\times 10^{-5} $ & $    $ & $    $ & $     $ & $  $ \\ [3pt]
159 & $ 00:54:46.21 $ & $ -37:47:18.6 $ & $   1.68 $ & $ 1.45\times 10^{1} $ & $   8.42 \pm   3.97\times 10^{0} $ & $   9.87 \pm   4.78\times 10^{-5} $ & $    $ & $    $ & $     $ & $  $ \\ [3pt]
160 & $ 00:54:56.43 $ & $ -37:39:38.1 $ & $   2.03 $ & $ 1.19\times 10^{1} $ & $   4.42 \pm   3.09\times 10^{0} $ & $   1.33 \pm   0.83\times 10^{-4} $ & $    $ & $    $ & $     $ & $  $ \\ [3pt]
161 & $ 00:54:41.54 $ & $ -37:43:03.9 $ & $   2.18 $ & $ 1.35\times 10^{1} $ & $   3.55 \pm   3.28\times 10^{0} $ & $   8.06 \pm  11.10\times 10^{-5} $ & $    $ & $    $ & $     $ & $  $ \\ [3pt]
162 & $ 00:54:45.95 $ & $ -37:45:21.6 $ & $   1.54 $ & $ 1.27\times 10^{1} $ & $   1.98 \pm   0.54\times 10^{1} $ & $   1.45 \pm   0.52\times 10^{-4} $ & $    $ & $    $ & $     $ & $  $ \\ [3pt]
163 & $ 00:55:11.74 $ & $ -37:40:14.8 $ & $   1.69 $ & $ 1.08\times 10^{1} $ & $   2.71 \pm   2.30\times 10^{0} $ & $   7.65 \pm   5.27\times 10^{-5} $ & $    $ & $    $ & $     $ & $  $ \\ [3pt]

\hline
\end{longtable} 
\end{landscape}

\onecolumn
\begin{landscape}\scriptsize
\begin{longtable}{llllllllll}
\caption{Summary of the optical counterparts of X-ray sources found in
  NGC~300 (see text for details.)\label{tab:table2}}\\
\hline
ID & $\alpha_\text{J2000.0}$ & $\delta_\text{J2000.0} $ &  mag (B) & mag (V) & mag (R)  & comments \\
(1) & (2) & (3) & (4) & (5) & (6) & (7) ) \\
\\ 
\hline
\\
\endfirsthead
\multicolumn{9}{l}{\slshape Continued}   \\
\hline
\\
ID & $\alpha_\text{J2000.0}$ & $\delta_\text{J2000.0} $ &  mag (B) & mag (V) & mag (R)  & comments \\
(1) & (2) & (3) & (4) & (5) & (6) & (7) \\

\\
\hline
\\
\endhead

  1 (1) & $ 00:55:10.0    $ & $ -37:42:12.2   $ & $    23.44 $ & $    23.35 $ & $    22.62 $ &  X (RP2001 - P42), WR* (SCA2003 - 41)                                                                                   \\
  3 (1) & $ 00:55:10.9    $ & $ -37:48:34.2   $ & $    20.82 $ & $    20.40 $ & $    19.85 $ &  X (RP2001 - P58), radio (PFP2004), USNO: 0450-00323113                                                                 \\
  7 (1) & $ 00:54:50.3    $ & $ -37:38:49.5   $ & $    22.77 $ & $    21.53 $ & $    20.42 $ &  X (RP2001 - P32), radio (PFP2004)                                                                                      \\
  8     & $   $ & $   $ & $   $ & $   $ & $   $ &  X (XMMU J005510.7-373855), X (RP2001 - P33)                                                                            \\
 12     & $   $ & $   $ & $   $ & $   $ & $   $ &  SNR (BL97 - N300-S10), HII (SMJ96 - HII W22), HII (BL97 - N300-H10), radio (PFP2004)                                   \\
 13     & $   $ & $   $ & $   $ & $   $ & $   $ &  X (RP2001 - P25)                                                                                                       \\
 17     & $   $ & $   $ & $   $ & $   $ & $   $ &  X (RP2001 - P41), HII (SMJ96 - HII W7), HII (SMJ96 - HII W9), SNR (PDL2000 - SNR 6)                                    \\
 20     & $   $ & $   $ & $   $ & $   $ & $   $ &  SNR (PDL2000 - SNR 15), USNO: 0450-00324001, USNO: 0450-00324043                                                       \\
 21     & $   $ & $   $ & $   $ & $   $ & $   $ &  X (RP2001 - P50)                                                                                                       \\
 26 (1) & $ 00:55:26.2    $ & $ -37:38:37.9   $ & $    24.33 $ & $    23.69 $ & $    23.07 $ &  X (RP2001 - P30)                                                                                                       \\
 28 (1) & $ 00:54:42.6    $ & $ -37:43:43.2   $ & $    22.88 $ & $    22.60 $ & $    22.18 $ &  Cepheid (G84 - 14)                                                                                                     \\
 31 (1) & $ 00:54:30.6    $ & $ -37:43:15.4   $ & $    24.35 $ & $    23.82 $ & $    23.35 $ &                                                                                                                         \\
 34     & $   $ & $   $ & $   $ & $   $ & $   $ &  SNR (BL97 - N300-S26), HII (DCL88 - 141), radio (PFP2004)                                                              \\
 37     & $   $ & $   $ & $   $ & $   $ & $   $ &                                                                                                                         \\
 38 (1) & $ 00:54:42.6    $ & $ -37:37:32.7   $ & $    22.15 $ & $    21.71 $ & $    21.40 $ &                                                                                                                         \\
 39     & $   $ & $   $ & $   $ & $   $ & $   $ &  Assoc* (PGF2001 - AS 56a)                                                                                              \\
 40     & $   $ & $   $ & $   $ & $   $ & $   $ &  X (RP2001 - P26)                                                                                                       \\
 41     & $   $ & $   $ & $   $ & $   $ & $   $ &  X (RP2001 - P54)                                                                                                       \\
 42     & $   $ & $   $ & $   $ & $   $ & $   $ &  Cepheid (G84 - 17), X (RP2001 - P51)                                                                                   \\
 43     & $   $ & $   $ & $   $ & $   $ & $   $ &                                                                                                                         \\
 46 (1) & $ 00:55:42.9    $ & $ -37:44:35.5   $ & $    15.26 $ & $    15.35 $ & $    16.31 $ &  X (RP2001 - P48), USNO: 0450-00326180                                                                                  \\
 46 (2) & $ 00:55:42.9    $ & $ -37:44:35.5   $ & $    15.25 $ & $    15.30 $ & $    16.10 $ &                                                                                                                         \\
 47     & $   $ & $   $ & $   $ & $   $ & $   $ &                                                                                                                         \\
 48     & $   $ & $   $ & $   $ & $   $ & $   $ &                                                                                                                         \\
 50     & $   $ & $   $ & $   $ & $   $ & $   $ &                                                                                                                         \\
 52 (1) & $ 00:55:42.2    $ & $ -37:40:24.7   $ & $    24.81 $ & $    23.41 $ & $    22.14 $ &  X (RP2001 - P36)                                                                                                       \\
 54     & $   $ & $   $ & $   $ & $   $ & $   $ &  radio (PFP2004)                                                                                                        \\
 56     & $   $ & $   $ & $   $ & $   $ & $   $ &  HII (SMJ96 - HII C27), radio (PFP2004)                                                                                 \\
 57     & $   $ & $   $ & $   $ & $   $ & $   $ &                                                                                                                         \\
 60 (1) & $ 00:54:44.4    $ & $ -37:41:15.8   $ & $    23.04 $ & $    22.31 $ & $    21.52 $ &  SNR (PDL2000 - SNR 5), X (RP2001 - P41)                                                                                \\
 64     & $   $ & $   $ & $   $ & $   $ & $   $ &                                                                                                                         \\
 65     & $   $ & $   $ & $   $ & $   $ & $   $ &                                                                                                                         \\
 66     & $   $ & $   $ & $   $ & $   $ & $   $ &  X (RP2001 - P50)                                                                                                       \\
 67 (1) & $ 00:54:35.9    $ & $ -37:34:33.9   $ & $    24.48 $ & $    23.49 $ & $    22.47 $ &                                                                                                                         \\
 69     & $   $ & $   $ & $   $ & $   $ & $   $ &  radio (PFP2004)                                                                                                        \\
 71     & $   $ & $   $ & $   $ & $   $ & $   $ &  USNO: 0450-00323152                                                                                                    \\
 72 (1) & $ 00:54:37.6    $ & $ -37:42:49.0   $ & $    24.98 $ & $    24.35 $ & $    23.62 $ &  SNR (PDL2000 - SNR 3)                                                                                                  \\
 73     & $   $ & $   $ & $   $ & $   $ & $   $ &  X (RP2001 - P28)                                                                                                       \\
 74     & $   $ & $   $ & $   $ & $   $ & $   $ &                                                                                                                         \\
 79 (1) & $ 00:54:22.2    $ & $ -37:40:25.3   $ & $    21.78 $ & $    21.73 $ & $    21.91 $ &  HII (DCL88 - 10), SNR? (BL97 - N300-S2), USNO: 0450-00318453                                                           \\
 84     & $   $ & $   $ & $   $ & $   $ & $   $ &                                                                                                                         \\
 87     & $   $ & $   $ & $   $ & $   $ & $   $ &                                                                                                                         \\
 88     & $   $ & $   $ & $   $ & $   $ & $   $ &  Star (HG86 - 12), USNO: 0450-00318469                                                                                  \\
 90     & $   $ & $   $ & $   $ & $   $ & $   $ &                                                                                                                         \\
 91     & $   $ & $   $ & $   $ & $   $ & $   $ &  radio (PFP2004)                                                                                                        \\
 92     & $   $ & $   $ & $   $ & $   $ & $   $ &                                                                                                                         \\
 94 (1) & $ 00:55:12.2    $ & $ -37:38:23.8   $ & $    24.00 $ & $    23.58 $ & $    23.11 $ &                                                                                                                         \\
 99     & $   $ & $   $ & $   $ & $   $ & $   $ &  X (RP2001 - P57)                                                                                                       \\
100     & $   $ & $   $ & $   $ & $   $ & $   $ &                                                                                                                         \\
102     & $   $ & $   $ & $   $ & $   $ & $   $ &  HII (BL97 - N300-H3), Assoc* (PGF2001 - AS 14)                                                                         \\
103     & $   $ & $   $ & $   $ & $   $ & $   $ &                                                                                                                         \\
107     & $   $ & $   $ & $   $ & $   $ & $   $ &  USNO: 0450-00321929                                                                                                    \\
112     & $   $ & $   $ & $   $ & $   $ & $   $ &                                                                                                                         \\
117 (1) & $ 00:55:13.9    $ & $ -37:47:57.4   $ & $    24.64 $ & $    24.28 $ & $    23.43 $ &                                                                                                                         \\
120 (1) & $ 00:54:53.2    $ & $ -37:43:11.3   $ & $    20.98 $ & $    20.15 $ & $    19.35 $ &  radio (PFP2004)                                                                                                        \\
122     & $   $ & $   $ & $   $ & $   $ & $   $ &                                                                                                                         \\
123     & $   $ & $   $ & $   $ & $   $ & $   $ &  SNR? (BL97 - N300-S19)                                                                                                 \\
125     & $   $ & $   $ & $   $ & $   $ & $   $ &                                                                                                                         \\
126 (1) & $ 00:54:53.3    $ & $ -37:40:29.0   $ & $    23.32 $ & $    23.03 $ & $    22.59 $ &                                                                                                                         \\
128     & $   $ & $   $ & $   $ & $   $ & $   $ &  USNO: 0450-00324361                                                                                                    \\
132     & $   $ & $   $ & $   $ & $   $ & $   $ &                                                                                                                         \\
134     & $   $ & $   $ & $   $ & $   $ & $   $ &                                                                                                                         \\
135     & $   $ & $   $ & $   $ & $   $ & $   $ &                                                                                                                         \\
136 (1) & $ 00:54:44.8    $ & $ -37:37:40.9   $ & $    24.16 $ & $    23.81 $ & $    23.53 $ &                                                                                                                         \\
137     & $   $ & $   $ & $   $ & $   $ & $   $ &                                                                                                                         \\
139 (1) & $ 00:55: 7.6    $ & $ -37:45:15.5   $ & $    24.20 $ & $    24.26 $ & $    24.29 $ &                                                                                                                         \\
140     & $   $ & $   $ & $   $ & $   $ & $   $ &  X (RP2001 - P30)                                                                                                       \\
141     & $   $ & $   $ & $   $ & $   $ & $   $ &                                                                                                                         \\
142 (1) & $ 00:54:33.2    $ & $ -37:44: 4.4   $ & $    24.80 $ & $    24.33 $ & $    23.99 $ &                                                                                                                         \\
143     & $   $ & $   $ & $   $ & $   $ & $   $ &                                                                                                                         \\
145 (1) & $ 00:55: 1.0    $ & $ -37:34:40.5   $ & $    19.87 $ & $    18.50 $ & $    17.22 $ &  USNO: 0450-00322130                                                                                                    \\
146     & $   $ & $   $ & $   $ & $   $ & $   $ &                                                                                                                         \\
147 (1) & $ 00:54:51.8    $ & $ -37:47: 8.6   $ & $    24.87 $ & $    24.25 $ & $    23.26 $ &                                                                                                                         \\
148     & $   $ & $   $ & $   $ & $   $ & $   $ &                                                                                                                         \\
150     & $   $ & $   $ & $   $ & $   $ & $   $ &                                                                                                                         \\
151 (1) & $ 00:55:33.9    $ & $ -37:43:12.8   $ & $    21.28 $ & $    21.07 $ & $    20.71 $ &  SNR? (BL97 - N300-S28), HII (S66b - 80), HII (DCL88 - 159), Star (BGK2002 - A14), radio (PFP2004), USNO: 0450-00325259 \\
151 (2) & $ 00:55:33.8    $ & $ -37:43: 9.3   $ & $    24.15 $ & $    24.10 $ & $    24.28 $ &                                                                                                                         \\
152     & $   $ & $   $ & $   $ & $   $ & $   $ &                                                                                                                         \\
153     & $   $ & $   $ & $   $ & $   $ & $   $ &                                                                                                                         \\
155 (1) & $ 00:54:56.8    $ & $ -37:43:41.6   $ & $    24.83 $ & $    24.04 $ & $    23.40 $ &                                                                                                                         \\
155 (2) & $ 00:54:56.9    $ & $ -37:43:38.7   $ & $    24.27 $ & $    24.14 $ & $    24.44 $ &                                                                                                                         \\
156 (1) & $ 00:54:30.0    $ & $ -37:40:31.2   $ & $    24.83 $ & $    23.95 $ & $    23.48 $ &                                                                                                                         \\
157 (1) & $ 00:54:52.7    $ & $ -37:46: 2.1   $ & $    25.67 $ & $    24.78 $ & $    23.65 $ &                                                                                                                         \\
158 (1) & $ 00:54:53.4    $ & $ -37:44:40.7   $ & $    21.66 $ & $    21.56 $ & $    21.54 $ &  USNO: 0450-00321373                                                                                                    \\
158 (2) & $ 00:54:53.3    $ & $ -37:44:39.2   $ & $    23.27 $ & $    22.45 $ & $    21.58 $ &                                                                                                                         \\
159 (1) & $ 00:54:46.2    $ & $ -37:47:19.0   $ & $    23.61 $ & $    22.73 $ & $    21.47 $ &                                                                                                                         \\
160     & $   $ & $   $ & $   $ & $   $ & $   $ &  HII (DCL88 - 99), radio (PFP2004)                                                                                      \\
161 (1) & $ 00:54:41.5    $ & $ -37:43: 3.9   $ & $    22.89 $ & $    21.24 $ & $    20.07 $ &                                                                                                                         \\
162 (1) & $ 00:54:46.0    $ & $ -37:45:21.1   $ & $    23.90 $ & $    22.31 $ & $    21.14 $ &                                                                                                                         \\
163 (1) & $ 00:55:11.6    $ & $ -37:40:13.8   $ & $    23.23 $ & $    23.02 $ & $    23.22 $ &                                                                                                                         \\

\hline
\end{longtable}
\end{landscape}

\end{document}